\documentclass[11pt,a4paper,iop,apj,twocolappendix]{emulateapj}
\usepackage{epsfig}
\begin{document}
\title{Testing systematics of Gaia DR2 parallaxes with empirical surface brightness - color relations applied to eclipsing binaries}
\author{Dariusz Graczyk\altaffilmark{1,2,3},  Grzegorz Pietrzy{\'n}ski\altaffilmark{3,2},Wolfgang Gieren\altaffilmark{2,1}, Jesper Storm\altaffilmark{5}, \\ Nicolas Nardetto\altaffilmark{6}, 
Alexandre Gallenne\altaffilmark{6,7}, Pierre F. L. Maxted\altaffilmark{8}, Pierre Kervella\altaffilmark{9,10}, \\ Zbigniew Ko{\l}aczkowski\altaffilmark{3,11}, Piotr Konorski\altaffilmark{4}, Bogumi{\l} Pilecki\altaffilmark{3}, Bart{\l}omiej Zgirski\altaffilmark{3}, Marek G{\'o}rski\altaffilmark{2}, \\Ksenia Suchomska\altaffilmark{3}, Paulina Karczmarek\altaffilmark{1,2}, M{\'o}nica Taormina\altaffilmark{3}, Piotr Wielg{\'o}rski\altaffilmark{3}, Weronika Narloch\altaffilmark{1,2,3}, \\ Rados{\l}aw Smolec\altaffilmark{3}, Rolf Chini\altaffilmark{12,13} and Louise Breuval\altaffilmark{10}} 
\affil{$^1$Millennium Institute of Astrophysics (MAS), Chile}
\affil{$^2$Universidad de Concepci{\'o}n, Departamento de Astronomia, Casilla 160-C, Concepci{\'o}n, Chile; darek@astro-udec.cl}
\affil{$^3$Centrum Astronomiczne im. Miko{\l}aja Kopernika (CAMK), PAN, Bartycka 18, 00-716 Warsaw, Poland; darek@ncac.torun.pl}
\affil{$^4$Obserwatorium Astronomiczne, Uniwersytet Warszawski, Al.~Ujazdowskie 4, 00-478, Warsaw, Poland}
\affil{$^5$Leibniz-Institut f\"{u}r Astrophysik Potsdam, An der Sternwarte 16, 14482 Potsdam, Germany}
\affil{$^6$Universit\'e C$\hat{\rm o}$te d'Azur, Observatoire de la C$\hat{\rm o}$te d'Azur, CNRS, Laboratoire Lagrange, UMR7293, Nice, France}
\affil{$^7$European Southern Observatory, Alonso de C{\'o}rdova 3107, Casilla 19001, Santiago 19, Chile}
\affil{$^8$Astrophysics Group, Keele University, Staffordshire, ST5 5BG, UK}
\affil{$^9$Unidad Mixta Internacional Franco-Chilena de Astronom{\'i}a (CNRS UMI 3386), Departamento de Astronom{\'i}a,
Universidad de Chile, Camino El Observatorio 1515, Las Condes, Santiago, Chile}
\affil{$^{10}$LESIA (UMR 8109), Observatoire de Paris, PSL Research University, CNRS, UPMC, Univ. Paris-Diderot, 5 place Jules Janssen,
92195 Meudon, France}
\affil{$^{11}$Instytut Astronomiczny, Uniwersytet Wroc{\l}awski, Kopernika 11, 51-622 Wroc{\l}aw,  Poland}
\affil{$^{12}$Ruhr University Bochum, Faculty of Physics and Astronomy, Astronomical Institute, 44801 Bochum, Germany}
\affil{$^{13}$Instituto de Astronomia, Universidad Catolica del Norte Avenida Angamos 0610, Antofagasta, Chile}

\begin{abstract}
Using a sample of 81 galactic, detached eclipsing binary stars we investigated the global zero-point shift of their parallaxes with the {\it Gaia} Data Release 2 (DR2) parallaxes. The stars in the sample lay in a distance range of 0.04--2 kpc from the Sun. The photometric parallaxes $\varpi_{Phot}$ of the eclipsing binaries were determined by applying a number of empirical surface brightness - color (SBC) relations calibrated on optical-infrared colors. For each SBC relation we calculated the individual differences $ d\varpi_i=(\varpi_{Gaia}-\varpi_{Phot})_i$ and then we calculated unweighted and weighted means. As the sample covers the whole sky we interpret the weighted means as the global shifts of the {\it Gaia} DR2 parallaxes with respect to our eclipsing binary sample. Depending on the choice of the SBC relation the shifts vary from $-0.094$ mas to $-0.025$ mas. The weighted mean of the zero-point shift from all colors and calibrations used is $d\varpi=-0.054\pm0.024$ mas. However, the SBC relations based on $(B\!-\!K)$ and $(V\!-\!K)$ colors, which are the least reddening dependent and have the lowest intrinsic dispersions, give a zero-point shift of $d\varpi=-0.031\pm0.011$ mas in full agreement with results obtained by Lindegren et al. and Arenou et al.  Our result confirms the global shift of {\it Gaia} DR2 parallaxes of $d\varpi=-0.029$ mas reported by the {\it Gaia} team, but we do not confirm the larger zero-point shift reported by a number of follow-up papers.

\end{abstract} 

%---------------------------
\section{Introduction}
The {\it Gaia} mission \citep{gaia16} is a milestone in understanding the Milky Way structure and its chemo-dynamical evolution, but it is also extremely important for the recalibration of standard candles and standard rulers used to construct the extragalactic distance ladder. A number of studies devoted to the recalibration of Cepheids and RR Lyr stars with Gaia DR2 parallaxes has already appeared \citep[e.g.][]{groe18,mur18}. 

The possible global systematics of the {\it Gaia} DR2 parallaxes were evaluated with different methods by the Gaia consortium.  \cite{lind18} used quasars and internal validation solutions to estimate the systematics and reported that the parallaxes are, on average, by about 0.03 mas too small, and that some local significant correlations of parallaxes occur. \cite{are18} made a comparison of {\it Gaia} DR2 parallaxes with a number of external catalogs to evaluate the global shift. Individual shifts vary from $-0.118\pm0.003$ mas (HIPPARCOS) to $+0.09\pm0.07$ mas (Phoenix dwarf), where the minus sign signifies that {\it Gaia} parallaxes are too small. However, most catalogs point to a global systematic offset of about $-0.03$ mas, e.g. comparisons with the Large and Small Magellanic Cloud stars catalogs, which have the smallest formal uncertainties, yield shifts of $-0.0380\pm0.0004$ mas and $-0.0268\pm0.0006$ mas, respectively. 

The global shift was also evaluated by a number of external works. A comparison with distance moduli of 50 Galactic cepheids derived from HST photometry resulted in an estimate of global zero-point offset of {\it Gaia} DR2 parallaxes of $-0.046\pm0.013$ mas \citep{ries18}. Using a sample of 3475 red giant branch stars in the Kepler field from the APOKASC-2 catalog and asteroseismic relations, \cite{zinn18} found a systematic shift of $-0.053\pm2$({\it stat})$\pm1$({\it syst}) mas. The derived shift is specific to the {\it Kepler} field but it would also be a measure of the global zero-point shift. Another approach was presented by \cite{sta18} who used the eclipsing binary method. They calculated observed reddening-free bolometric fluxes $F_{bol}$ by fitting the spectral energy distribution (SED) and compared them to the bolometric luminosities $L_{bol}$ to obtain distances. Using a subsample of 89 suitable eclipsing binaries from their catalog \citep{sta16} they derived a global shift of $-0.082\pm0.033$ mas. The method used by \cite{sta18} is, from a theoretical point of view, the most robust one because it utilizes a very wide range of photometry from UV to midinfrared. However, the $L_{bol}$ are calculated from the effective temperature of each component and this radiative parameter is almost always the least constrained fundamental parameter of an eclipsing binary. There are two main reasons for this: 1) the wide range of methods used to determine effective temperatures that exhibit significantly different zero-points of the resulting temperature scales, and 2) the notorious difficulty in establishing an absolute temperature scale with a precision better than 1\% \cite[e.g.][]{cas14}. Thus from a practical point of view the method used by \cite{sta18} is not the optimal one because it has some systematics that are still difficult to properly evaluate. The need of an eclipsing binary approach which would minimize systematic effects, and would be based on direct and precise empirical relations, was the prime motivation for this work. As in our previous work  \citep[][hereafter G17]{gra17} we focus on an application of surface brightness - color relations.   

\section{The Sample}\label{sample}
We extended the sample of 35 eclipsing binaries compiled by G17 by searching for detached systems in the literature suitable for a precise distance determination.
In order to make the extension we used catalogs of eclipsing binary systems compiled by a number of authors: \cite{bil08,tor10,eke14,sou15,sta16}. 

\begin{figure}
\mbox{\includegraphics[width=0.44\textwidth]{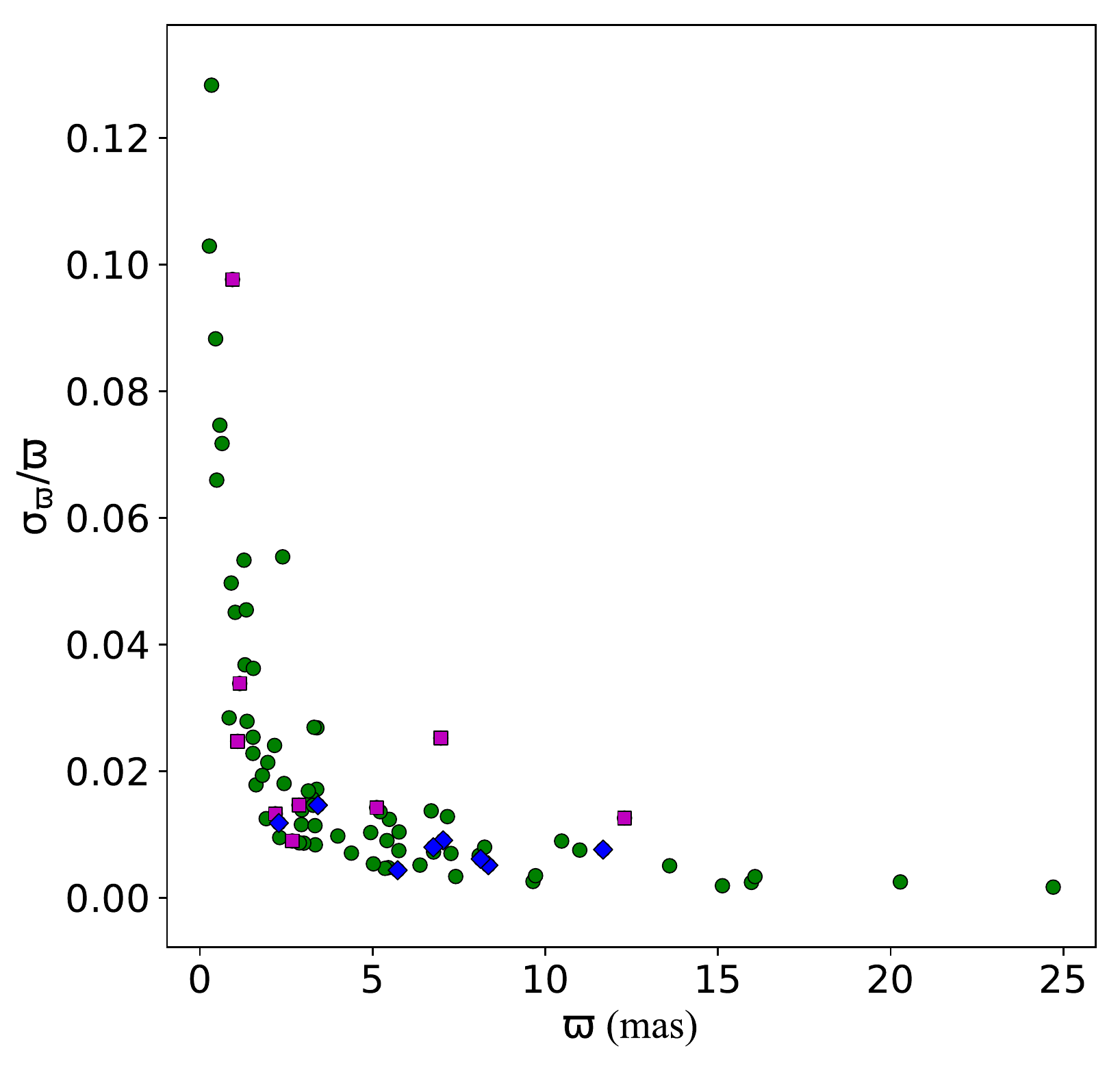}} 
\caption{The distribution of fractional error of {\it Gaia} DR2 parallaxes for all 81 eclipsing binaries in our sample. Nine systems with suspected binary motion detection or calibration problems are denoted by squares. Eight stars with
the significant PMa are denoted by diamonds.}
\label{fig:gaia_frac}
\end{figure}

We used the same selection criteria as in G17, however we relaxed the condition on a volume-limited sample ($d<300$ pc) by accepting also more distant systems. Because the $\beta$ Aur system is too bright to have a {\it Gaia} parallax, and AI Phe is a confirmed triple system (Konacki priv. communication) we drop these systems from our sample. We included also a system from our unpublished work: AL Dor.
Following G17 we retained systems with a fractional precision of recent {\it Gaia} DR2 parallaxes \citep{gaia16, gaia18, lind18} better then 10\%. 
A trigonometric parallax $\varpi$ with a fractional error $f=\sigma_\varpi/\varpi$ smaller than 0.1 is a good and weakly biased estimator of the true distance  \citep[e.g.][]{bai15,bai18}. In our sample only two early and distant systems have $f$ marginally larger than 0.1 i.e.: EM Car and DW Car. Most of the sample systems have fractional errors of their parallaxes smaller than 0.06 for which any bias can be completely neglected. Finally our sample contains 81 systems (51 on the northern hemisphere, 30 on the southern one). Their basic parameters are presented in Table~\ref{tab:sample} and the distribution of the parallax fractional errors is presented in Fig.~\ref{fig:gaia_frac}.

\cite{are18} proposed three quality controls to estimate reliability of astrometric solutions of {\it Gaia} DR2; they are defined by their equations (1--3). Those quality controls help to filter out most of spurious astrometric solutions, but they filter out also some well-defined solutions. In our sample 9 stars do not fulfill the first criterium (possible double stars, suspected binary motion or calibration problems) and 7 other stars do not fulfill the third criterium (i.e. {\verb visibility_periods_used }$\; <8$). However, the third criterium is useful mostly to very crowded fields in the Bulge and as no star in our sample is there, we kept these 7 systems during analysis. The 9 stars that do not fulfill the first criterium are denoted as squares in Fig.~\ref{fig:gaia_frac}. Only 3 of them have the relative parallax uncertainty larger by a factor of two than expected. We investigated our sample further by looking for the proper motion anomaly \citep[PMa;][]{ker19} which could be a sign of a photocenter movement or a third body in a system.  The PMa results from a comparison of the proper motions vectors at the {\it Hipparcos} and {\it Gaia} epoch with the mean proper motion computed from a coordinate shift between {\it Hipparcos} and {\it Gaia} epochs. We found a significant (SNR=5) detection of the PMa for 8 stars, but interestingly none of them are common with the subsample of 9 suspected stars. The stars with the PMa are denoted as diamonds in Fig.~\ref{fig:gaia_frac}. All of them have well defined 5-parametric astrometric solutions, and the PMa, if confirmed with later {\it Gaia} data releases, is likely due to long-period changes caused by an unrecognized triple companion.

\subsection{Photometry}

\subsubsection{Optical}
We used Tycho-2 $B_T$ and $V_T$
photometry \citep{hog00} downloaded from
Vizier \citep{och00}\footnote{http://vizier.u-strasbg.fr:
I/259/tyc2}. The Tycho photometry was
subsequently transformed onto the Johnson system using the method outlined by
\cite{bes00}. Whenever possible we used Johnson $B,V$
photometry from the compilation of \cite{mer97} and also absolute optical photometry from literature sources.
 
\subsubsection{Near infrared}
\label{sec:NIR}
We used NIR $JHK_S$
photometry of the Two Micron All Sky Survey (2MASS) \citep{skr06}
from Vizier\footnote{http://vizier.u-strasbg.fr:
II/281/2mass6x}. For the purpose of using the SBC relations based on the ($V\!-\!K$) color expressed on the Johnson photometric system we transformed the 2MASS magnitudes
using the equations given in \cite{bes88} and
\cite{car01}\footnote{\texttt{http://www.astro.caltech.edu/$\sim$jmc/2mass/v3/\\transformations/}}.
The transformation equations are as follows: 

{\small
\begin{eqnarray} 
\mkern-18mu K_J - K_{\rm 2M} & = & 0.037 - 0.017 (J-K)_{\rm 2M} - 0.007 (V-K)_{\rm 2M}  \\
\mkern-18mu (J-K)_J & = & 1.064 (J-K)_{\rm 2M} +0.006  \\ 
\mkern-18mu (H-K)_J & = &1.096 (H-K)_{\rm 2M} - 0.027 
\end{eqnarray}}

\begin{deluxetable*}{l@{}ccccccccc}
\tabletypesize{\scriptsize}
\tablecaption{Basic data on the selected detached eclipsing binaries \label{tab:sample}}
\tablewidth{0pt}
\tablehead{
\colhead{Name} & \colhead{GAIA DR2} & \colhead{RA$_{(2000)}$} & \colhead{DEC$_{(2000)}$} & \colhead{V} & \colhead{Spectral}&\colhead{Orbital}& \colhead{GAIA DR2 Parallax}\\
\colhead{} & \colhead{ID} & \colhead{h:m:s} & \colhead{deg:m:s} & \colhead{(mag)} & \colhead{Type}  & \colhead{Period (d)}  & \colhead{(mas)}}
\startdata
MU Cas & 429158427922077184 &00:15:51.560& +60:25:53.64&$\!\!\!$10.808$\pm$0.007 & B5V+B5V & $\!\!\!$9.652949& 0.453$\pm$0.040 \\
YZ Cas& 539047365205648128 & 00:45:39.077& +74:59:17.06& 5.653$\pm$0.015& A2m+F2V&4.4672235&$\!\!\!$10.473$\pm$0.094\\
V459 Cas$\;\;$& 522461335380031104 & 01:11:29.913 &+61:08:48.07 &$\!\!\!$10.322$\pm$0.003&A1V+A1V & 8.4582538 & 1.304$\pm$0.048\\ 
V505 Per& 455772347387763840 & 02:21:12.964& +54:30:36.28& 6.889$\pm$0.016& F5V+F5V&$\!\!\!$4.222020 &$\!\!\!$15.970$\pm$0.039\\
DN Cas&513431183821175936 & 02:23:11.540& +60:49:50.18& 9.878$\pm$0.010& O8V+B0V &$\;\;$2.31095111 &0.485$\pm$0.032
\enddata
\tablecomments{Only first five entries are shown, this Table is entirely avalaible in electronic form.}
\end{deluxetable*}

%\section{Photometry}

\section{Method}\label{method}
\subsection{Collection of fundamental parameters}
\label{sec:coll}
For each system we collected orbital
and photometric parameters from the literature including the most recent publications. 
We were searching for basic parameters describing 
dynamical and geometrical parameters of each system:
the radial velocity semi-amplitudes $K_{1,2}$,
the orbital period $P$, the orbital inclination $i$, the eccentricity $e$, the longitude of
periastron $\omega$, the photometric relative radii $r_{1,2}$. These parameters were supported by radiative parameters:
the effective temperatures $T_{1,2}$ of both components. The radiative parameters are usually known with the least precision and accuracy. Whenever several papers independently presented the analysis of a given eclipsing binary we adopted the weighted mean for the parameters. However, if there was a significant improvement on the precision of parameters reported in one of papers, then we used only values from that paper.  
It turned out that we could not always directly trace all the above parameters from literature sources. In some cases when modeling of an eclipsing binary was done with numerical codes based on the Roche formalism
(the Wilson-Devinney code, the ELC code, etc.) we had to calculate relative radii and radial velocity semiamplitudes from the absolute dimensions presented in the relevant papers. The collected parameters are summarized in Table~\ref{tab:par}.
 
\subsection{The Wilson-Devinney model of the systems}
\label{sec:wd}
For the purpose of obtaining homogenous parameters for the 
eclipsing binary sample we decided to create a model of each
system following G17. The models were built using the Wilson-Devinney (WD) code version
2007 \citep{wil71,wil79,wil90,van07} while parameters of the models were
based on solutions published in the literature (see Sec.~\ref{sec:coll}). 
All models were checked for internal
consistency of the parameters and it turned out that in many cases they had
to be fine-tuned. In particular, the temperature ratio and the absolute 
temperature scale, being important for a precise prediction of infrared 
light ratios, were inspected carefully.

The procedure was as follows.
Dynamical and geometrical parameters were transformed
into the semi-major axis of the system $a$, the mass ratio $q$ and into
dimensionless Roche potentials $\Omega_{1,2}$ using equations given in
\cite{tor10} and \cite{wil79}. Both $\Omega$ and $q$ are parameters directly fitted or used
within the WD program. We usually fixed the temperature of the primary star
$T_1$ and then, using published light ratios in different photometric
bands, we adjusted the temperature of the companion $T_2$. In very few cases,
however, we also rederived $T_1$ 
%\textcolor{red}
{using two temperature - color calibrations \citep{flo96,wor11}. 
However the temperature shifts are small and within errors given in literarure.} 
None of the eclipsing binaries in our sample has infrared $J,H,K$ light curves suitable for
deriving direct light ratios in those bands. Thus, in order to calculate
intrinsic infrared colors of the components of each system we employed
eclipsing binary models based on optical light curves and we extrapolated
them into the infrared. The extrapolation was done using the atmosphere approximation 
within the WD code which uses precomputed intensities based on Kurucz's ATLAS9 models \citep{kur93}.

The rotation parameter $F_{1,2}$ was kept to 1 (synchronous rotation), unless there
was a direct spectroscopic determination of $F$ significantly different
from unity. The albedo $A$ and the gravity brightening $g$ were set
in a standard way for a convective atmosphere cooler then 7200~K and
radiative ones for a hotter surface temperature. This was done only for
the sake of consistency because the two parameters have negligible effect on
the light ratios. The input and derived parameters used to create 
the appropriate WD models are listed in Table~\ref{tab:par}. 

\subsection{Correction of 2MASS magnitudes taken during eclipses}
Nine systems from our sample have 2MASS observations taken during the 
eclipses. To account for the light lost during the minima we used our WD models to
calculate the appropriate corrections which are given in Table~\ref{tab:2Mcorr}.

%\clearpage
\begin{turnpage}
\begin{deluxetable*}{@{}lcccccccccccccc@{}}
\tabletypesize{\scriptsize}
\tablecaption{Parameters of the Wilson-Devinney models \label{tab:par}}
\tablewidth{0pt}
\tablehead{
\colhead{}& \multicolumn{9}{c}{Input parameters}&\colhead{Ref.}&\multicolumn{4}{c}{WD model parameters}\\
\colhead{Eclipsing} & \multicolumn{2}{c}{RV semiamplitude}& \multicolumn{3}{c}{Orientation of the orbit}& \multicolumn{2}{c}{Fractional radius}& \multicolumn{2}{c}{Effective temperature}& \colhead{} & \colhead{Semimajor} & \colhead{Mass} & \colhead{$\Omega_1$} &\colhead{$\Omega_2$}\\
\colhead{binary} & \colhead{$K_1 ($km s$^{-1}$)}& \colhead{$K_2 ($km s$^{-1}$)} & \colhead{$e$} &\colhead{$\omega$ (rad)}& \colhead{$i$ (deg)} & \colhead{$r_1$} & \colhead{$r_2$} & \colhead{$T_1 ($K)} & \colhead{$T_2 ($K)}& \colhead{} & \colhead{axis$(R_\odot)$} & \colhead{ratio}&\colhead{}&\colhead{}}
\startdata
   MU Cas &  107.7(1.0)&  105.8(9) & 0.1930(3) & 0.234(7) & 87.02(7)  & 0.0917(9) & 0.1048(10) & 15900(500) &  15525(500)      &  1& 40.0234 & 1.0180 & 12.173& 10.952\\
    YZ Cas &  73.05(19)&  124.78(27) & 0.0 & --                   & 88.33(7)  & 0.14456(56) & 0.07622(33) & 9520(120) &  6880(240)     &  2& 17.4753 & 0.5854 & 7.5141 & 8.8912\\
V459 Cas &  81.70(60)&  83.90(60) & 0.0243(4) & 4.19(1) & 89.47(1)   & 0.07260(30) & 0.07100(30) & 9140(300) &  9085(300)     &  3& 27.6786 & 0.9738 & 14.776 & 14.757\\
  V505 Per &  89.01(8)&  90.28(9) & 0.0 & --                       & 87.95(4)   & 0.0860(9) & 0.0846(9) & 6512(21) &  6460(30)                 &  4& 14.9715 & 0.9859 & 12.618& 12.665\\
    DN Cas &  211(3)&  292(5) & 0.0 & --                              & 77.20(20) & 0.3070(20) & 0.2460(20) & 32100(1000) &  28500(1100) &  5& 23.5613 & 0.7226 & 4.0372&4.1129 
  \enddata 
\tablecomments{All references: 1 - \cite{slac04b}; 2 - \cite{pav14}; 3 - \cite{slac04a}; 4 - \cite{tom08a}; 5 - \cite{bak16}; 6 - this paper; 7 - \cite{cla01}; 8 - \cite{sou11} ; 9 - \cite{tom08b}; 10 - \cite{and91b}; 11 - \cite{gal16}; 12 - \cite{slac06}; 13 - \cite{mun04}; 14 - \cite{sou05a}; 15 - \cite{gro07}; 16 - \cite{dav16}; 17 - \cite{lac85}; 18 - \cite{max15}; 19 - \cite{slac02}; 20 - \cite{rib99}; 21 - \cite{imb02}; 22 - \cite{cla10}; 23 - \cite{and89}; 24 - \cite{hel09}; 25 - \cite{and87a}; 26 - \cite{tor14}; 27 - \cite{sab11}; 28 - \cite{kha01}; 29 - \cite{tom06}; 30 - \cite{sou05}; 31 - \cite{kir18}; 32 - \cite{tor15}; 33 - \cite{slac08}; 34 - \cite{pop85}; 35 - \cite{slac00}; 36 - \cite{lac87}; 37 - \cite{wil04}; 38 - \cite{hen06}; 39 - \cite{cla08}; 40 - \cite{sow12}; 41 - \cite{san18}; 42 - \cite{yak07}; 43 - \cite{alb14}; 44 - \cite{bak08}; 45 - \cite{and84}; 46 - \cite{and83}; 47 - \cite{and75b}; 48 - \cite{giu80};  49 - \cite{sou07b}; 50 - \cite{and89b}; 51 - \cite{sti94}; 52 - \cite{rat10}; 53 - \cite{cla07}; 54 - \cite{alb13}; 55 - \cite{slac11}; 56 - \cite{and75}; 57 - \cite{gro77}; 58 - \cite{wal83}; 59 - \cite{and84b}; 60 - \cite{lat96}; 61 - \cite{lac97}; 62 - \cite{pop98}; 63 - \cite{and93}; 64 - \cite{bud15}; 65 - \cite{slac12}; 66 - \cite{nor97}, 67 - \cite{and85}; 68 - \cite{fek11}; 69 - \cite{pri18}; 70 - \cite{tor09b}; 71 - \cite{slac14}; 72 - \cite{suc15}; 73 - \cite{pop71}; 74 - \cite{cla86}; 75 - \cite{imb85}; 76 - \cite{slac12b}; 77 - \cite{pop82}; 78 - \cite{alb09}; 79 - \cite{pop86}; 80 - \cite{imb86}; 81 - \cite{slac89}; 82 - \cite{ver15}; 83 - \cite{san16}; 84 - \cite{and87b}; 85 - \cite{alb07}; 86 - \cite{tor17}; 87 - \cite{pav09}; 88 - \cite{tka14}; 89 - \cite{hel15}; 90 - \cite{raw16}; 91 - \cite{tor10};  92 - \cite{lac87b}; 93 - \cite{har14}; 94 - \cite{tor00}; 95 - \cite{tor09};  96 - \cite{imb87};  97 - \cite{cla91}; 98 - \cite{les18}, 99 - \cite{pop87}; 100 - \cite{gri13}; 101 - \cite{sou13}; 102 - \cite{gra16}; 103 - \cite{tor99}; 104 - \cite{vos12}; 105 - \cite{bau95}; 106 - \cite{slac04c}, 107 - \cite{pop83};  108 - \cite{dem94}; 109 - \cite{cla10b}; 110 - \cite{slac14b}; 111 - \cite{cak09}\\
Only first five entries are shown, this Table is entirely avalaible in electronic form.} 
\end{deluxetable*}
\end{turnpage}

\begin{deluxetable}{@{}lccc@{}}
\tabletypesize{\scriptsize}
\tablecaption{Corrections to original 2MASS magnitudes \label{tab:2Mcorr}}
\tablewidth{0pt}
\tablehead{
\colhead{ID} & \colhead{K (mag)} &\colhead{H (mag)} & \colhead{J (mag)}}
\startdata
GG Lup & -0.345 & -0.342 & -0.334 \\
KX Cnc & -0.426 &-0.427 &-0.429 \\
WZ Oph & -0.535 &-0.536 &-0.540 \\
YZ Cas & -0.162 &-0.155 &-0.137 \\
BW Aqr & -0.524 &-0.526 &-0.532 \\
SZ Cen & -0.454 &-0.458 &-0.467 \\
V442 Cyg &-0.283&-0.283&-0.283 \\
ASAS1800 &-0.512&-0.505 &-0.493 \\
DW Car&-0.464&-0.465 &-0.467
\enddata
\end{deluxetable}

\subsection{Reddening} 
\label{sec:red} 
\begin{figure}
\mbox{\includegraphics[width=0.4\textwidth]{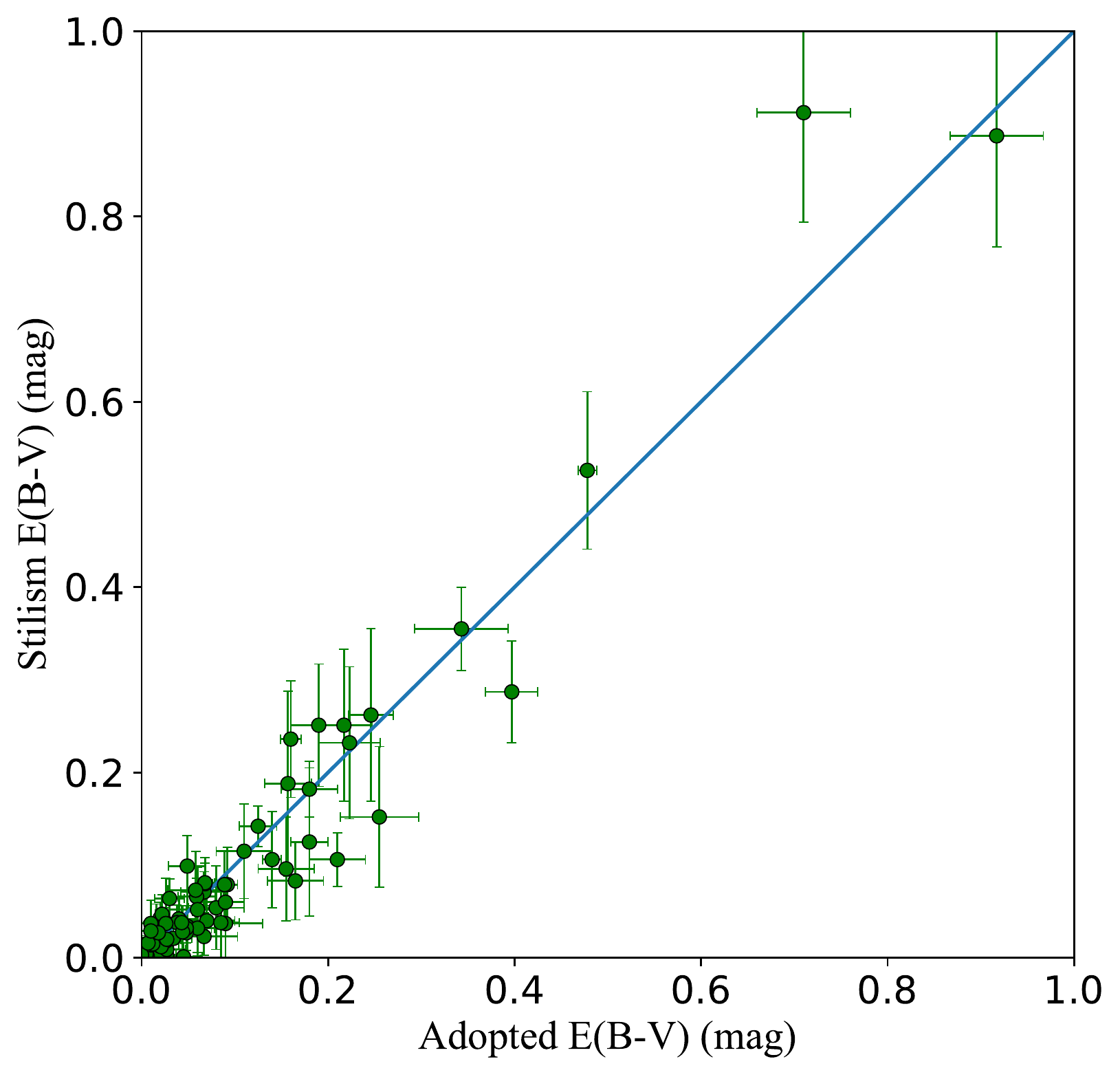}} 
\caption{The comparison of our adopted reddening estimates with 3D {\it Stilism} extinctions.}
\label{fig:ebv}
\end{figure}

Reddenings to each object were taken from the literature. We also derived independently values of $E(B\!-\!V)$
using the extinction maps of \cite{sch98} following the prescription given in \cite{suc15}. Usually as a final value of the extinction we
used an average, unless determinations were discrepant or we have at our disposal only one reddening estimate. The adopted reddenings are reported in Tab.~\ref{tab:fot}. To check them we compared them with independent
extinction estimates from 3D extinction maps from {\it Stilism} \citep{lal14,cap17} -- see Fig.~\ref{fig:ebv}. On average {\it Stilism} extinctions are slightly smaller, but an overall agreement is good, and although the spread is quite large both sets of reddenings are consistent within errors.  

%To re-derive temperatures we used a number of calibrations
%given below: 
%\begin{itemize}
%\item{$b-y$: \cite{hol07}, \cite{ram05}, \cite{alo96}, \cite{nap93}.}
%\item{$B\!-\!V$: \cite{cas10}, \cite{gon09}, \cite{ram05}, \cite{flo96}.}
%\item{$V\!-\!J$: \cite{cas10}, \cite{gon09}.}
%\item{$V\!-\!K$: \cite{wor11}, \cite{cas10}, \cite{gon09}, \cite{mas06}, \cite{ram05}, \cite{hou00}, \cite{alo96}.}
%\end{itemize}

%\subsection{Radial velocity semi-amplitudes}
%\label{sec:rv}
%Usually we assumed radial velocity semi-amplitudes from the literature. When
%two or more orbital solutions were published based on different radial
%velocity sets and having uncertainties of the same order of magnitude,
%we used the weighted mean to derive the final parameters. In few
%cases we redetermined the spectroscopic orbits from source data in order
%to derive directly $K_{1,2}$ or to check the consistency of the orbital
%parameters and their errors (G17). The spectroscopic orbits were derived with
%the Wilson-Devinney code taking into account the full model of a system and
%all proximity effects. A set of numerical constants used to change from SI
%units into astrophysical units were chosen after \cite{tor10}. 

%\clearpage
\begin{turnpage}
\begin{deluxetable*}{@{}lccccccccccccc@{}}
\tabletypesize{\scriptsize}
%\tabletypesize{\tiny}
\tablecaption{Physical and photometric parameters of individual components used to derive photometric parallaxes. \label{tab:fot}}
\tablewidth{0pt}
\tablehead{
\colhead{Eclipsing} & \colhead{E$(B-V)$ } & \multicolumn{2}{c}{Radius ($R_\odot$)} &\multicolumn{5}{c}{Observed magnitudes} & \multicolumn{5}{c}{Light ratios} \\ 
\colhead{binary} & \colhead{mag} & \colhead{$R_1$} & \colhead{$R_2$} &\colhead{$B$ (mag)} & \colhead{$V$ (mag)} &\colhead{$J$ (mag)} & \colhead{$H$ (mag)} & \colhead{$K$ (mag)} & \colhead{$B$} & \colhead{$V$} & \colhead{$J$} & \colhead{$H$}& \colhead{$K$}}
\startdata
     MU Cas &  0.478(10)& 3.670(43) & 4.194(48) & 11.112(9) &  10.808(7) &  10.170(23) & 10.092(23) &  10.050(16) & 1.253 & 1.257 & 1.273 & 1.276 & 1.277  \\
    YZ Cas &  0.015(10)& 2.526(11) & 1.332(6) & 5.715(26) &  5.660(15) &  5.490(20) & 5.502(42) &  5.475(22) & 0.061 & 0.088 & 0.168 & 0.200 & 0.205  \\
  V459 Cas &  0.246(24)& 2.009(13) & 1.965(13) & 10.591(9) &  10.322(3) &  9.791(23) & 9.715(25) &  9.668(16) & 0.936 & 0.941 & 0.949 & 0.950 & 0.951  \\
  V505 Per &  0.003(5)& 1.288(14) & 1.267(14) & 7.269(30) &  6.846(20) &  6.120(70) & 5.793(39) &  5.795(21) & 0.924 & 0.935 & 0.952 & 0.958 & 0.959  \\
    DN Cas &  0.917(50)& 7.233(96) & 5.796(82) & 10.495(18) &  9.878(10) &  8.413(25) & 8.217(19) &  8.132(24) & 0.492 & 0.500 & 0.521 & 0.527 & 0.529  \enddata                                                                 
\tablecomments{Only first five entries are shown, this Table is entirely avalaible in electronic form.}                                                     
\end{deluxetable*}
\end{turnpage}

\subsection{Intrinsic magnitudes}
\label{sec:intrin}
In Table~\ref{tab:fot} we summarize all parameters used to derive the intrinsic
photometric indexes of the component stars. In order to calculate them the photometry was dereddened using the mean Galactic interstellar extinction curve from \cite{fit07} assuming $R_V=3.1$. Then the light ratios in the Johnson $BVJHK$ bands were
derived using the WD models following G17, and employed to calculate the individual magnitudes and colors. 
%\textcolor{red}
{Both the extinction correction and extrapolation of light ratios into the infrared add to uncertainty on derived intrinsic magnitudes. The extinction correction error is given in Table~\ref{tab:fot} but for the extrapolation uncertainty we did not add additional error. While for systems with similar temperature of components extrapolation leads to negligible additional error in some systems with the large temperature ratio of components this uncertainty may be significant. However when calculating a parallax (or a distance) to a particular eclipsing binary as an average from two components this uncertainty largely cancels out.}

\subsection{Photometric parallaxes}
The eclipsing binary method gives the photometric distance as well as the photometric parallax to a particular target, because both quantities are {\it inferred} from observables. 
Because of this we decided to work in a regime of parallaxes instead of distances to avoid an additional bias, which may arise in the conversion of parallaxes into distances. 
However, we underline here that working with a regime of parallaxes does not solve the whole problem because, for fractional parallax uncertainties larger then 15\%--20\% the trigonometric parallax becomes
a poor prior on the distance.  
 
\subsubsection{Parallaxes from the bolometric flux scaling}
\label{par:bol}
In order to directly compare our results with those of \cite{sta18} we employed the bolometric flux scaling method utilizing the $V$-band bolometric corrections $BC$ to derive
photometric parallaxes. The photometric parallax $\varpi$ to the $i$-th component of the system was calculated using
equation: 
\begin{equation}
%\varpi_{i} (\rm mas) = 3.360\cdot10^{-8} R_{i}\, T^2\!\!\!_{i}\; 10^{0.2(BC_{i} + V_{i})}, 
\varpi_{i} ({\rm mas}) = 2.956\cdot10^{10} R^{-1}_{i}\, T^{-2}_{i}\; 10^{-0.2(BC_{i} + V_{i})}
\label{eq:1}
\end{equation}
where index $i=\{1,2\}$, $R$ is the radius of a component expressed in solar radii,
$T$ is its effective temperature, $BC$ is a bolometric correction
interpolated from the \cite{flo96} tables for a given temperature and $V$
is the intrinsic magnitude of a component. The
parallax to a particular system was calculated as the unweighted average parallax
of the two components. 

%\begin{figure}
%\mbox{\includegraphics[width=0.44\textwidth]{par_diff_hist_gaia_V_VK_diBenedetto2005.pdf}} 
%\caption{The histogram of parallax differences ({\it Gaia} $-$ EB). The photometric parallaxes are calculated using the SBC relation by \cite{diB05}.}
%\label{fig:hist:Ben}
%\end{figure}
 
\subsubsection{Parallaxes from the SBC relations}
\label{par:sbc}
The eclipsing binary method of distance determination is very flexible and has many different approaches.
We decided to use as our main approach the method based on the empirical surface brightness - color relations. 
In order to derive the photometric parallax to an eclipsing binary we need to calculate the individual angular
diameters of the components. An angular diameter is calculated with the formula:
\begin{equation}
\phi ({\rm mas}) = 10^{0.2\cdot(S-m_0)}
\end{equation}
where $S$ is the surface brightness in a given band and $m_0$ is the dereddened magnitude of a star in that band. In most cases $S$ is calibrated as a function of an intrinsic color of a star, usually in the form of a polynomial. 
The photometric parallax $\varpi_{Phot}$ follows from the equation:
\begin{equation}
\varpi_{Phot} ({\rm mas}) = 107.52  \cdot  \phi ({\rm mas}) / R ({\mathcal{R}_\odot}),
\label{eq:2}
\end{equation} 
where $R$ is the radius of the star and the conversion factor is equal to $ 
{\rm 1AU}/2 \mathcal{R}_\odot $ assuming the solar nominal radius $\mathcal{R}_\odot=695700$ km \citep{hab08,mam15}
and a length of the astronomical unit $ {\rm 1AU} = 149597871$ km \citep{pit09}.

We determined parallaxes using several SBC relations from the literature calibrated on infrared colors \citep{ker04,diB05,boy14,cha14}. The most 
useful calibrations of $S$ are based on $(B\!-\!K)$ and $(V\!-\!K)$ colors, because the reddening vector is almost parallel to the SBC relations
for intermediate- and late-type stars. Moreover, these relations have very low intrinsic dispersion \citep[e.g.][]{ker04,gra17}. However, we also
used calibrations based on other infrared colors to have a more detailed view of possible systematics. The parallax to a given eclipsing binary was calculated as the unweighted average of the individual parallaxes 
of its components.

\begin{figure*}
\centering
\begin{minipage}[t]{0.32\textwidth}
\includegraphics[width=0.95\textwidth]{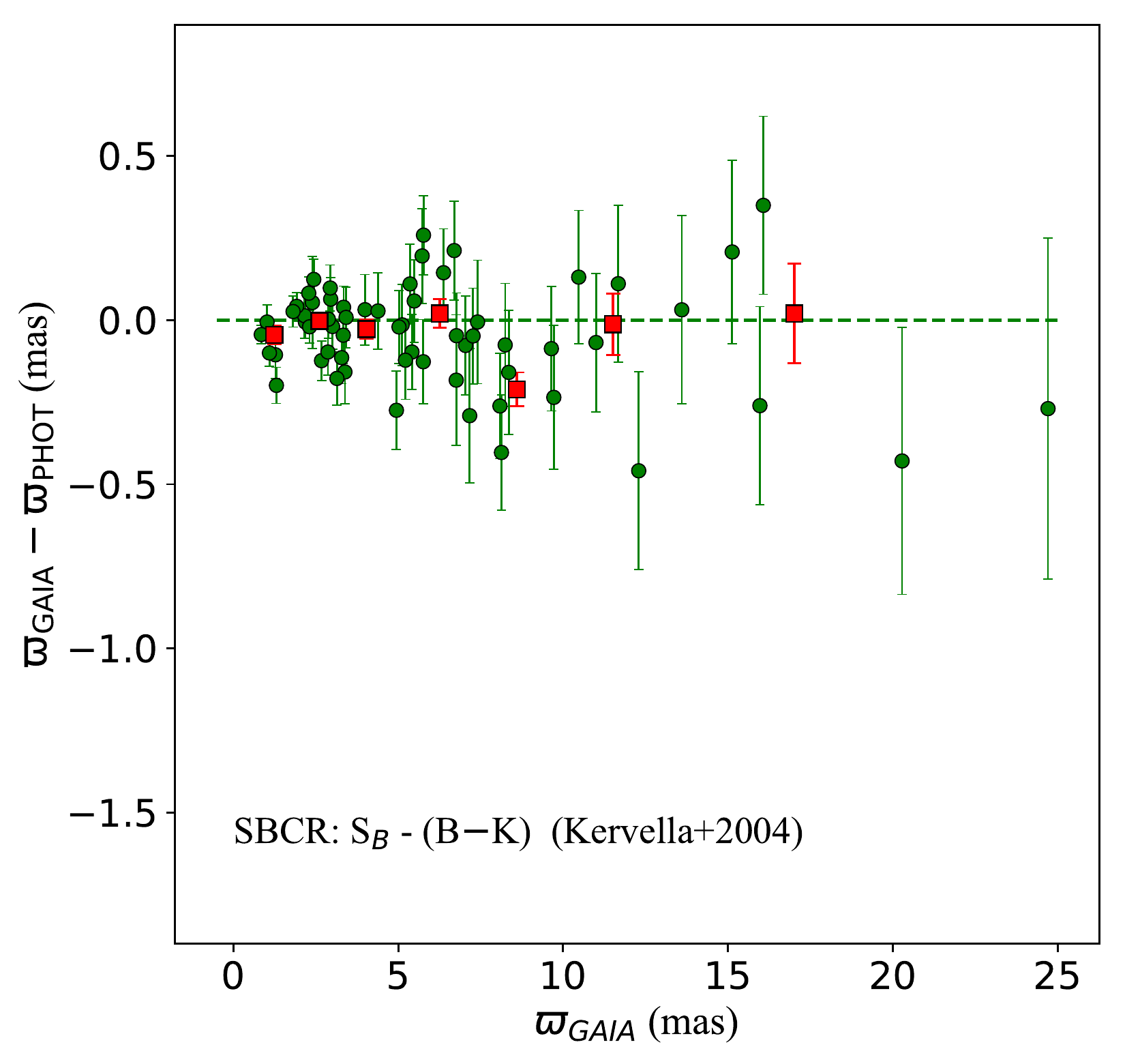} \\
\includegraphics[width=0.95\textwidth]{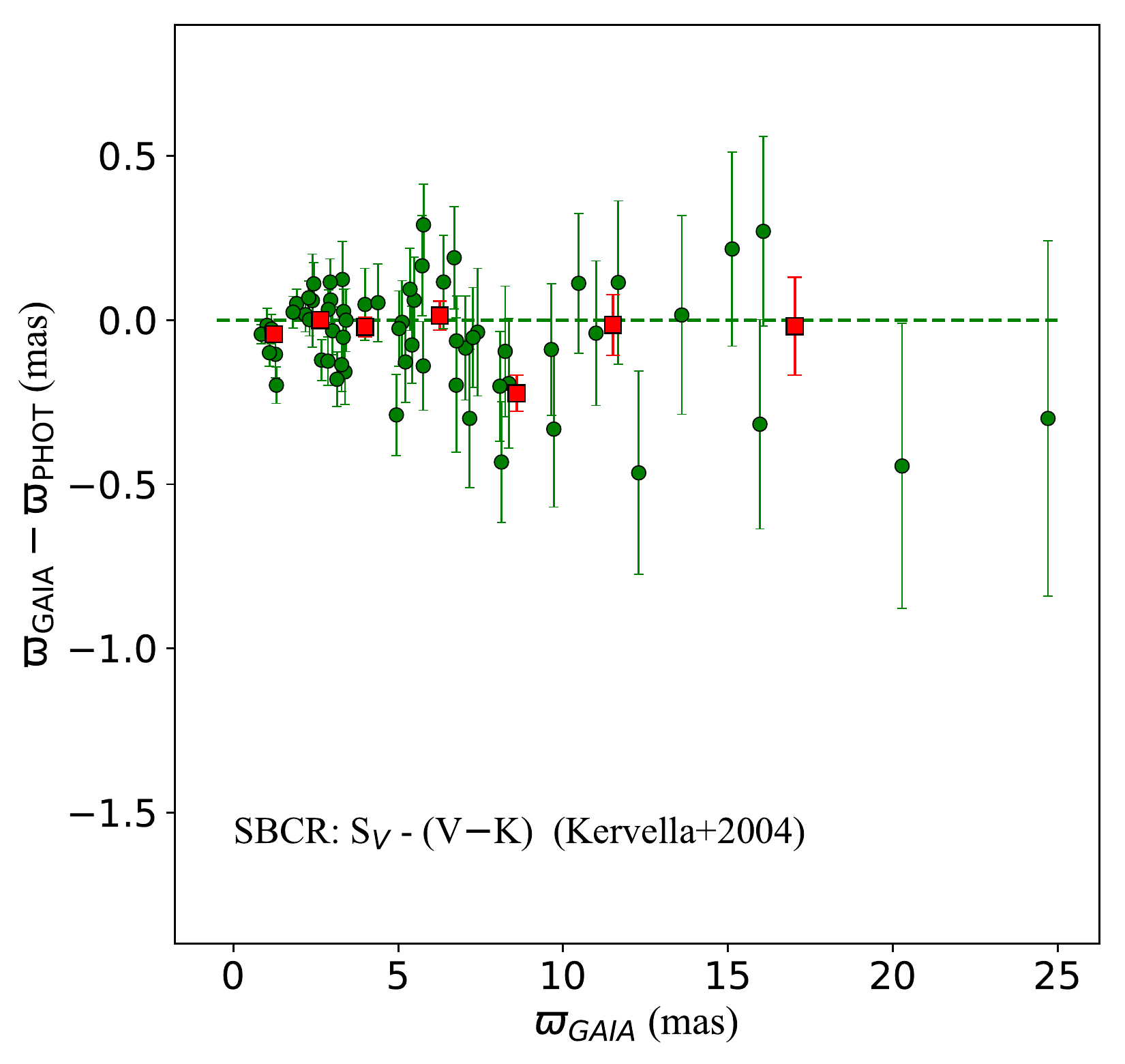} \\
\includegraphics[width=0.95\textwidth]{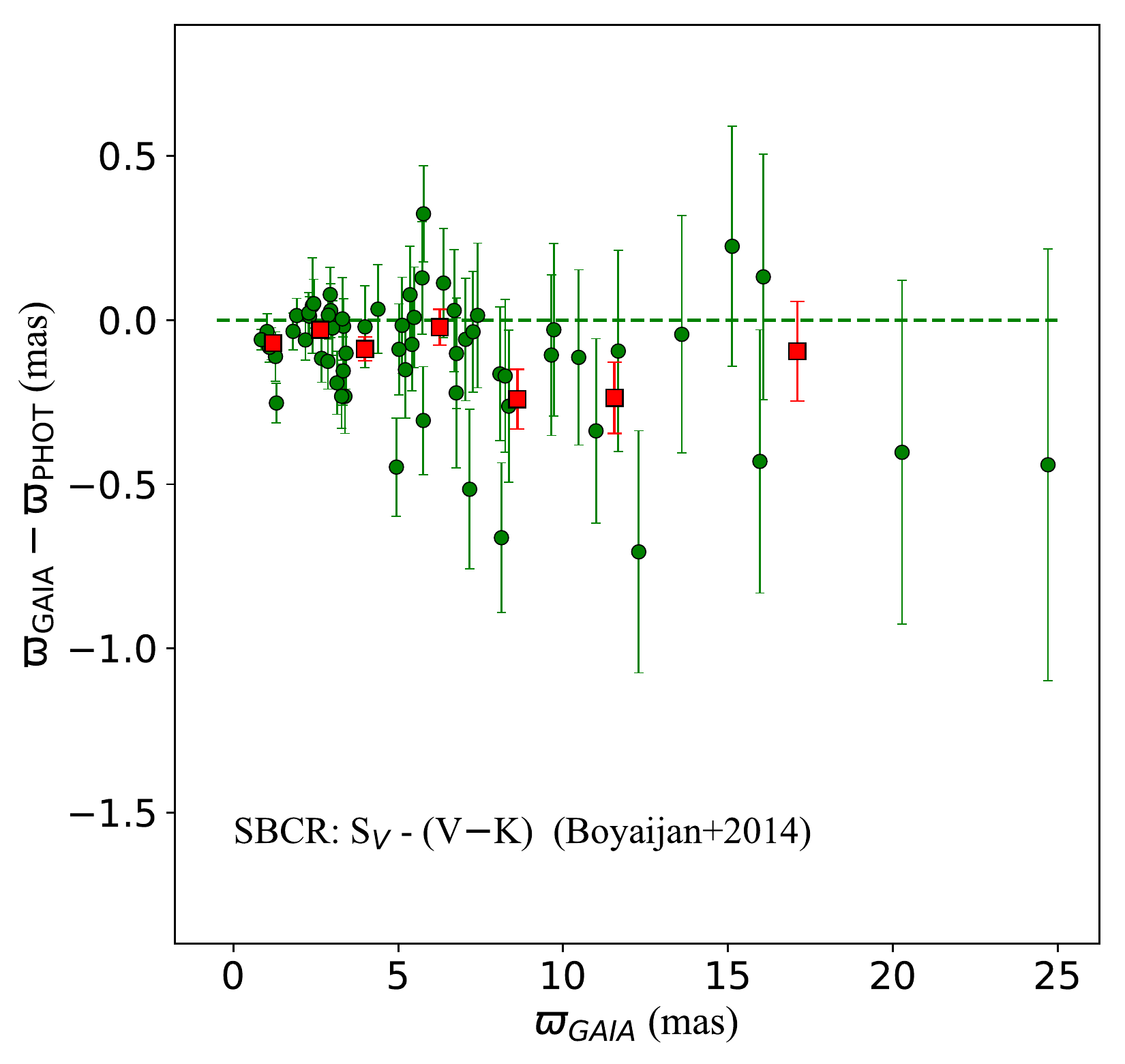} \\
\includegraphics[width=0.95\textwidth]{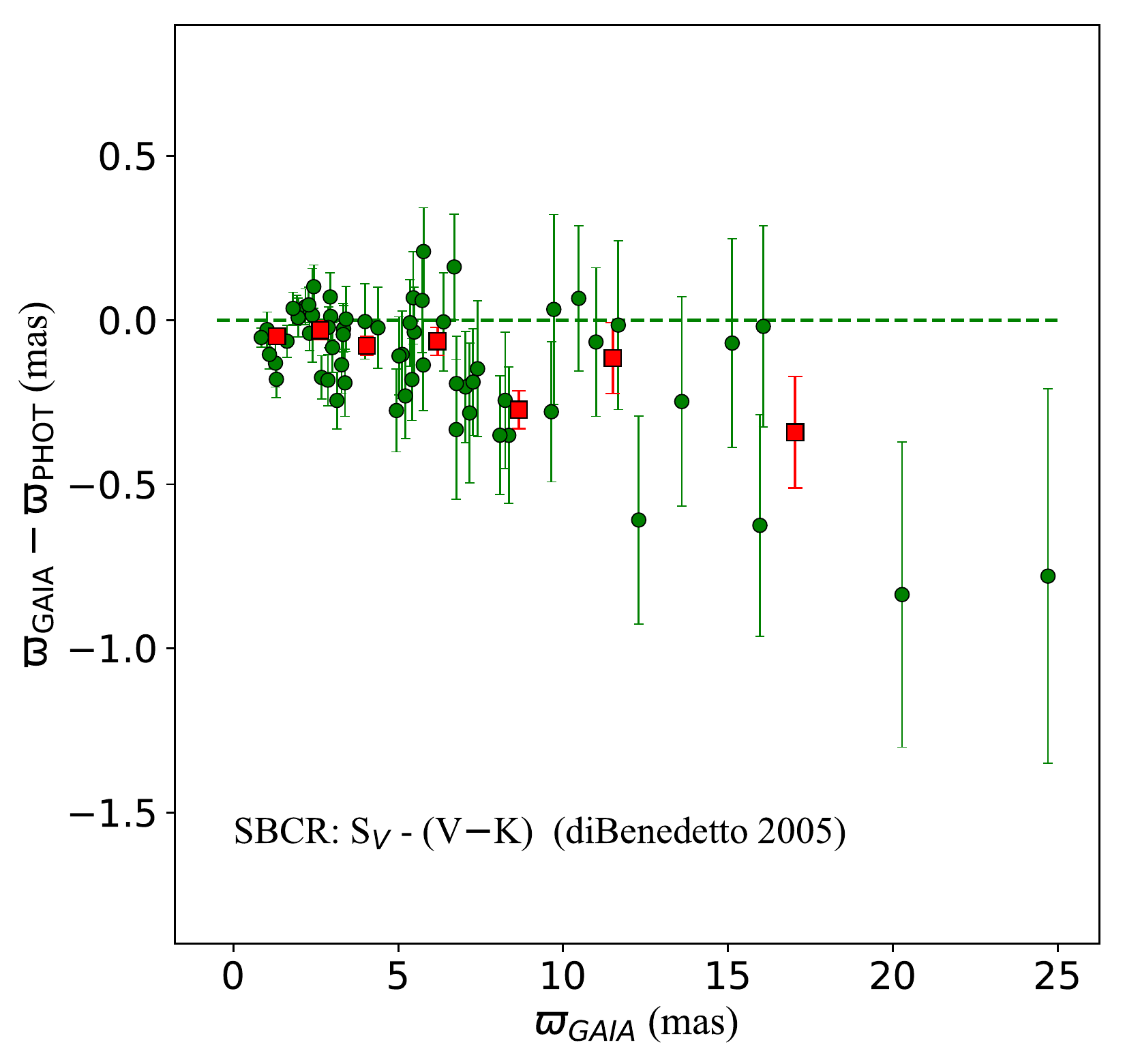}
\end{minipage}
\begin{minipage}[t]{0.32\textwidth}
\includegraphics[width=0.95\textwidth]{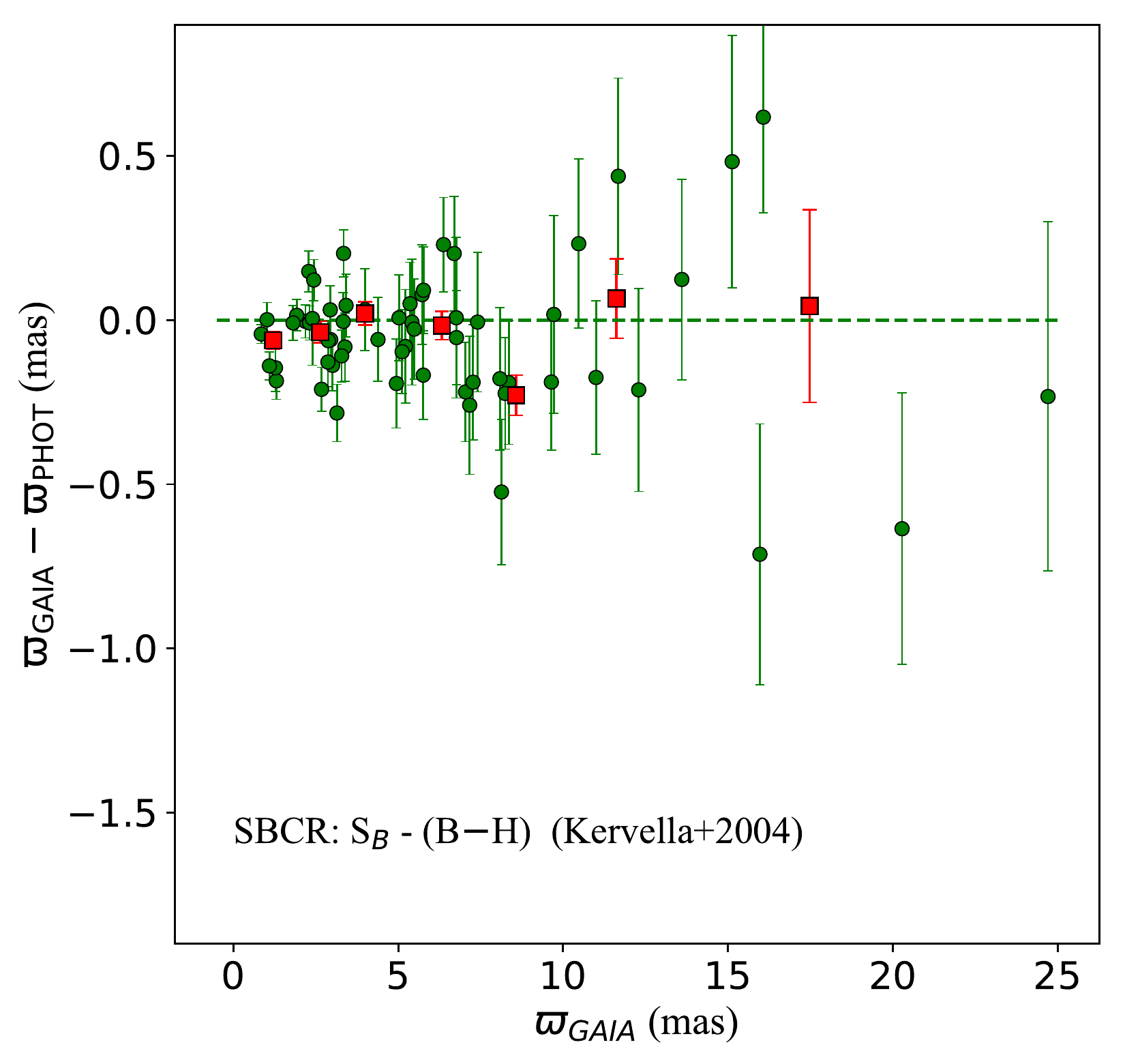}\\
\includegraphics[width=0.95\textwidth]{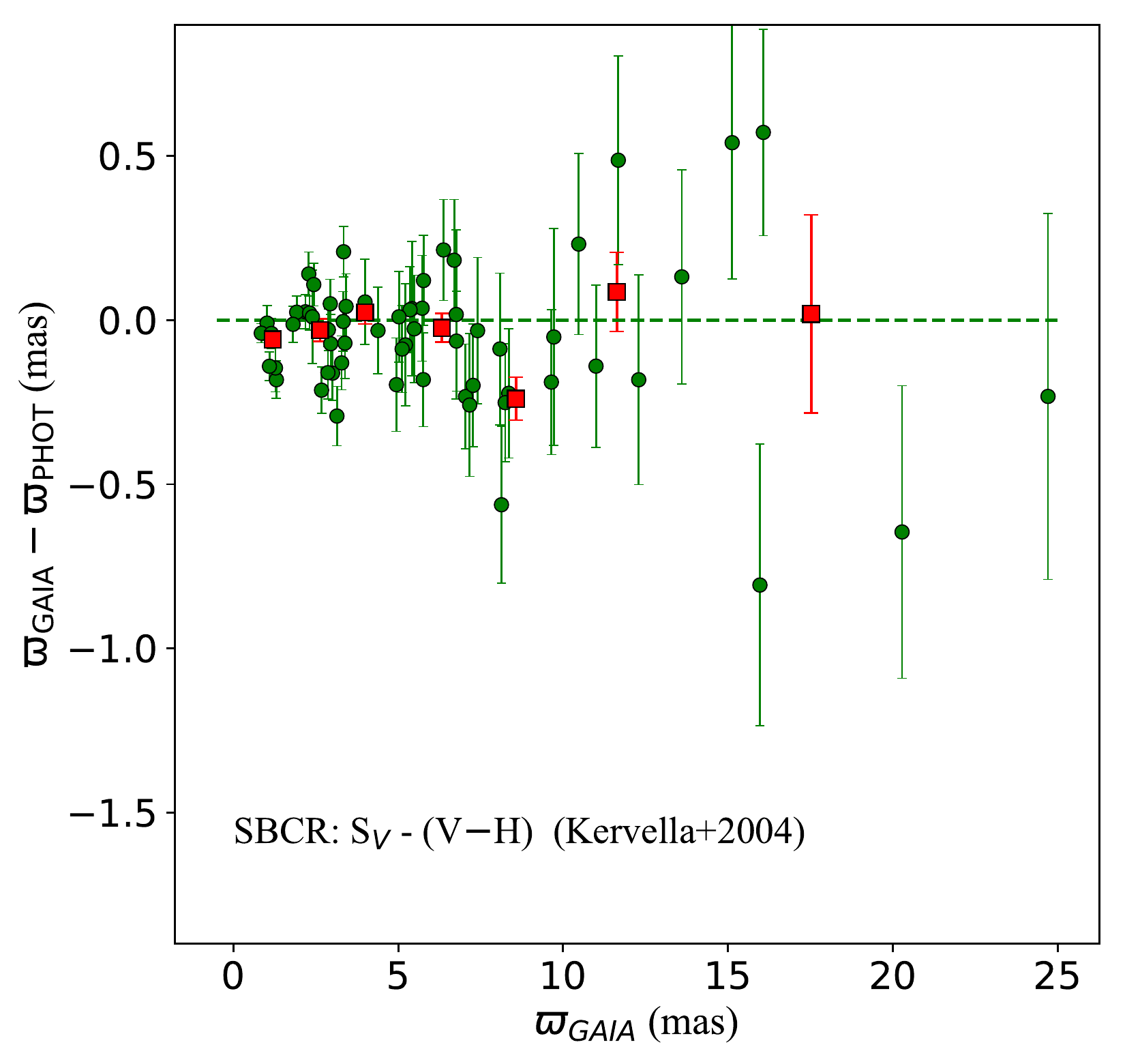}\\
\includegraphics[width=0.95\textwidth]{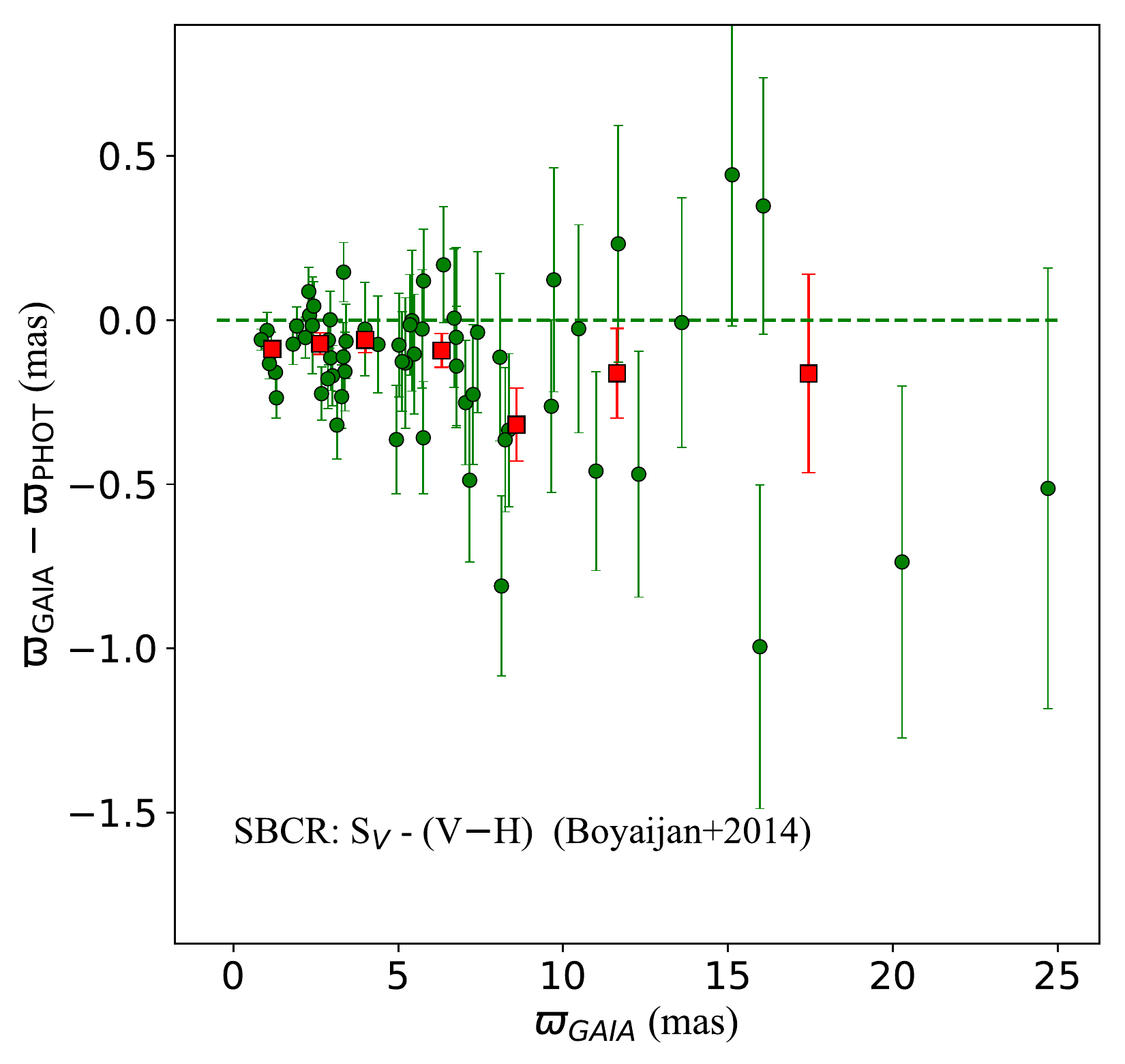}\\
\includegraphics[width=0.95\textwidth]{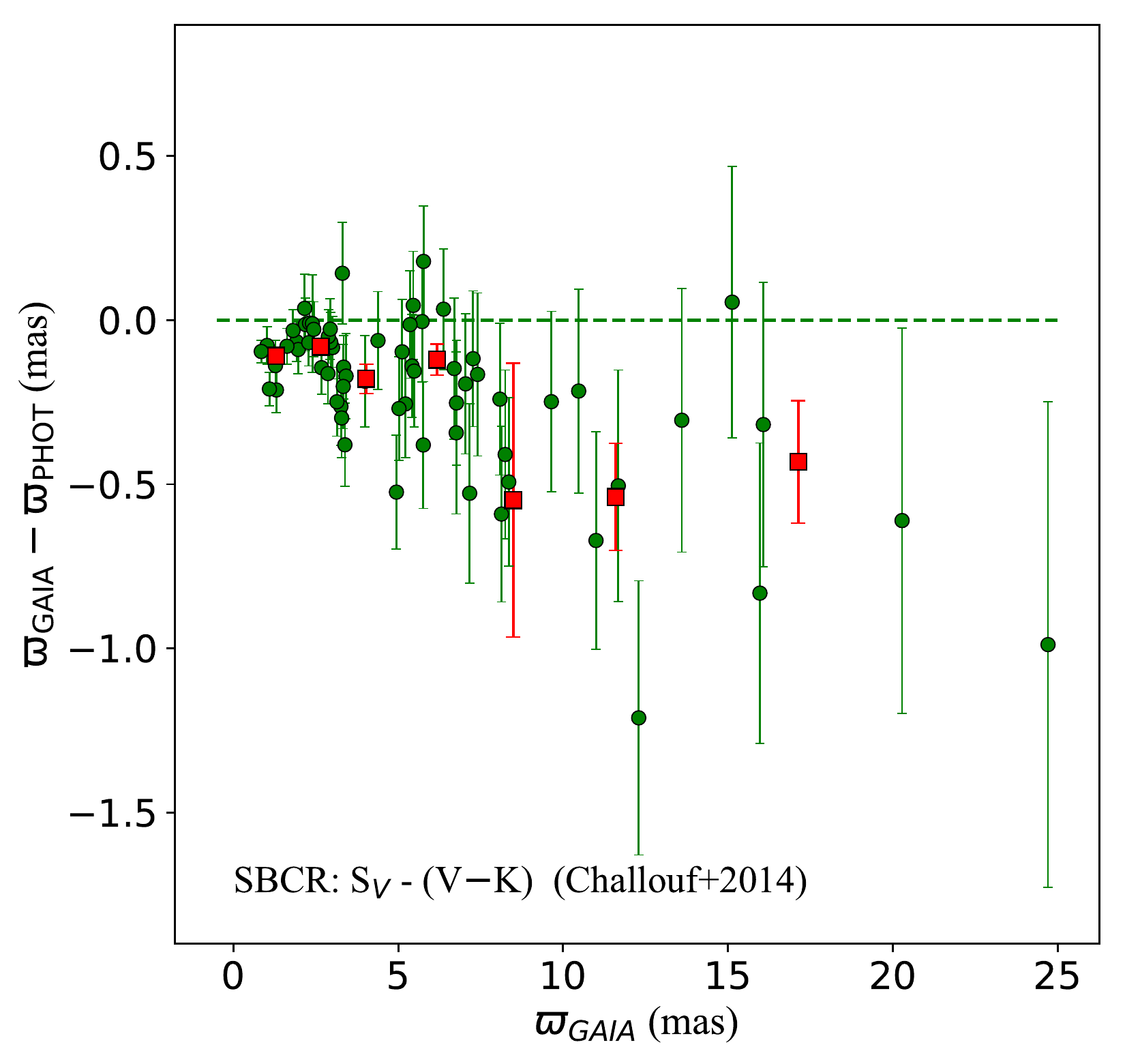}
\end{minipage}
\begin{minipage}[t]{0.32\textwidth}
\includegraphics[width=0.95\textwidth]{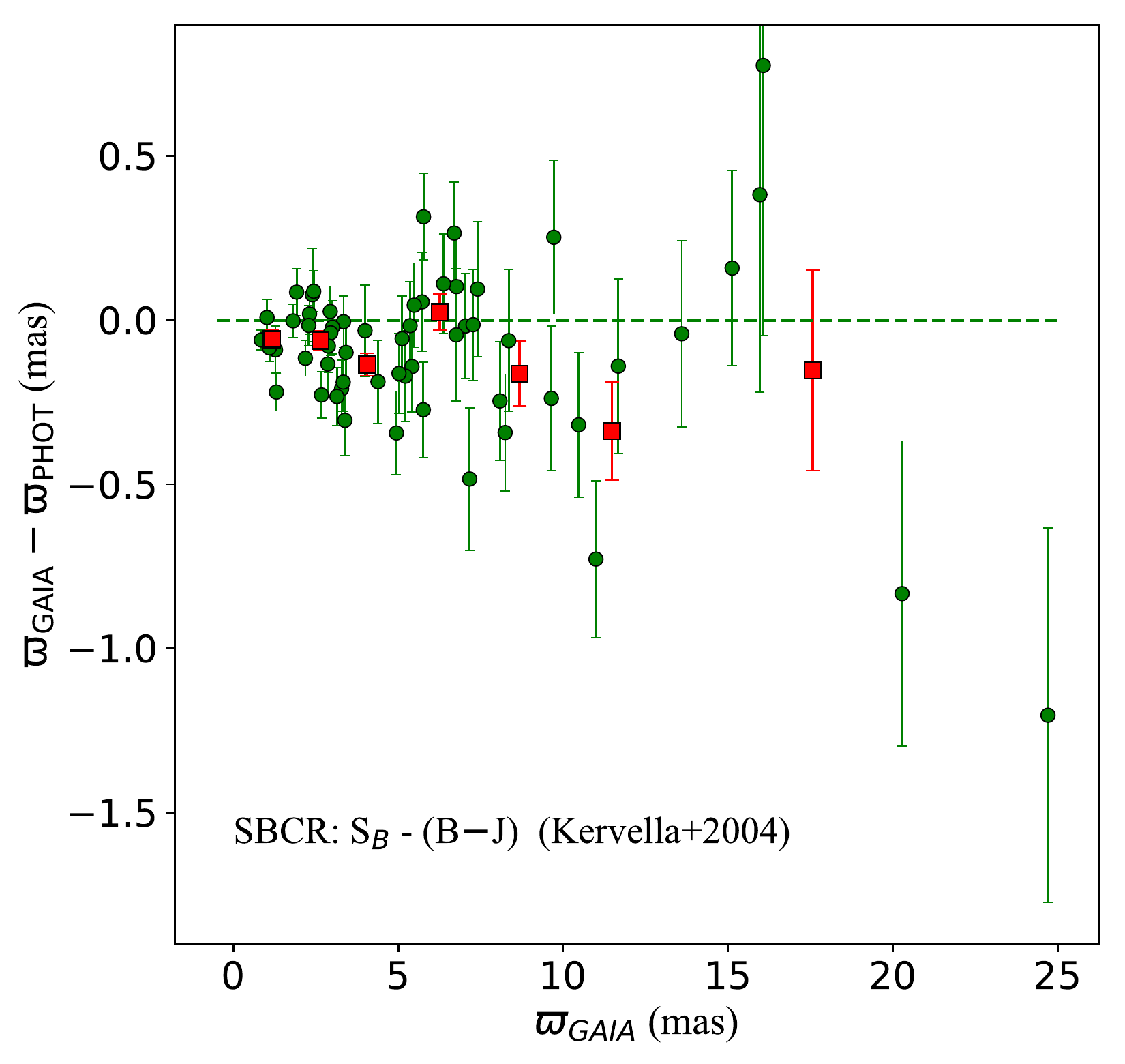}\\
\includegraphics[width=0.95\textwidth]{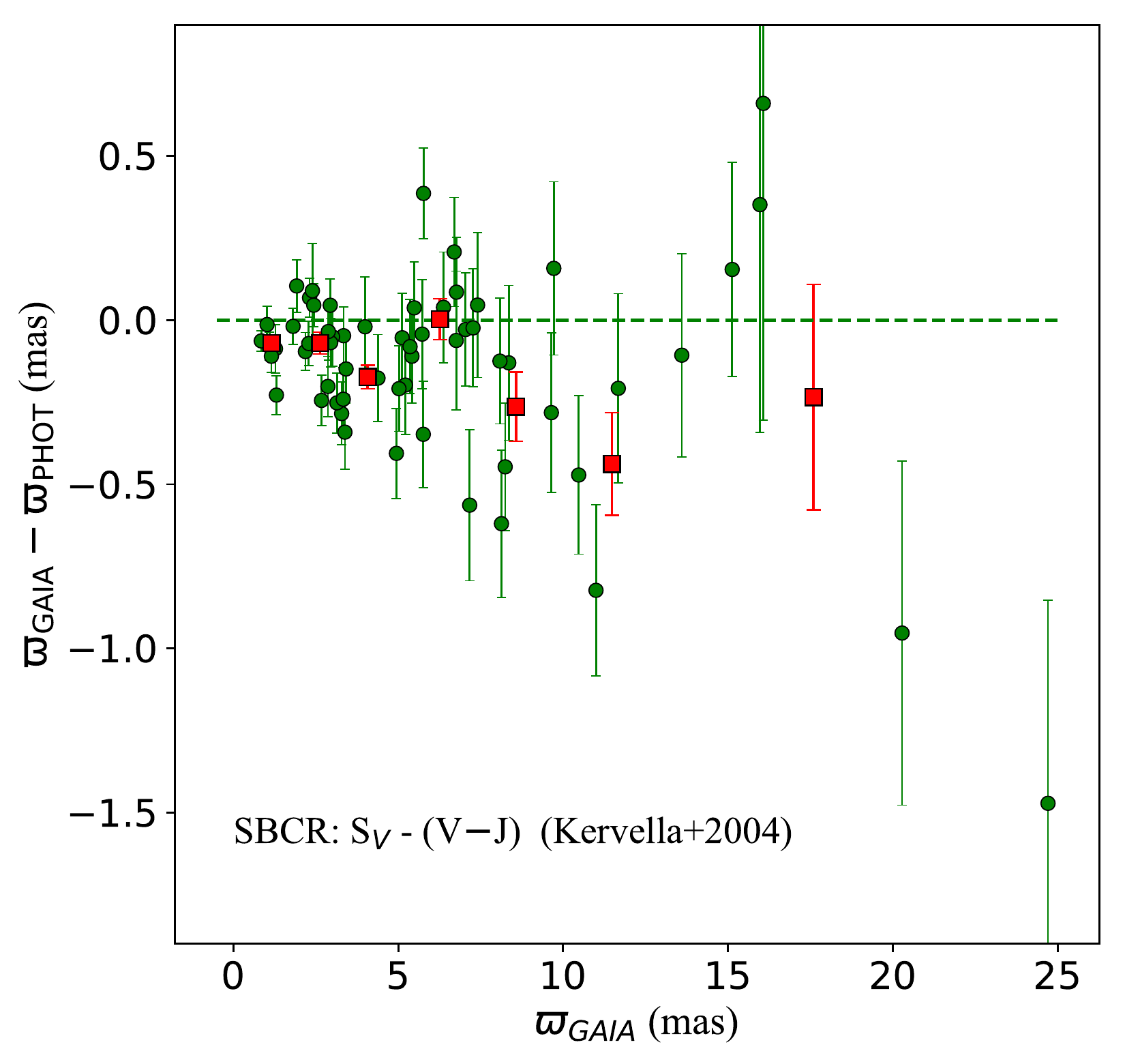}\\
\includegraphics[width=0.95\textwidth]{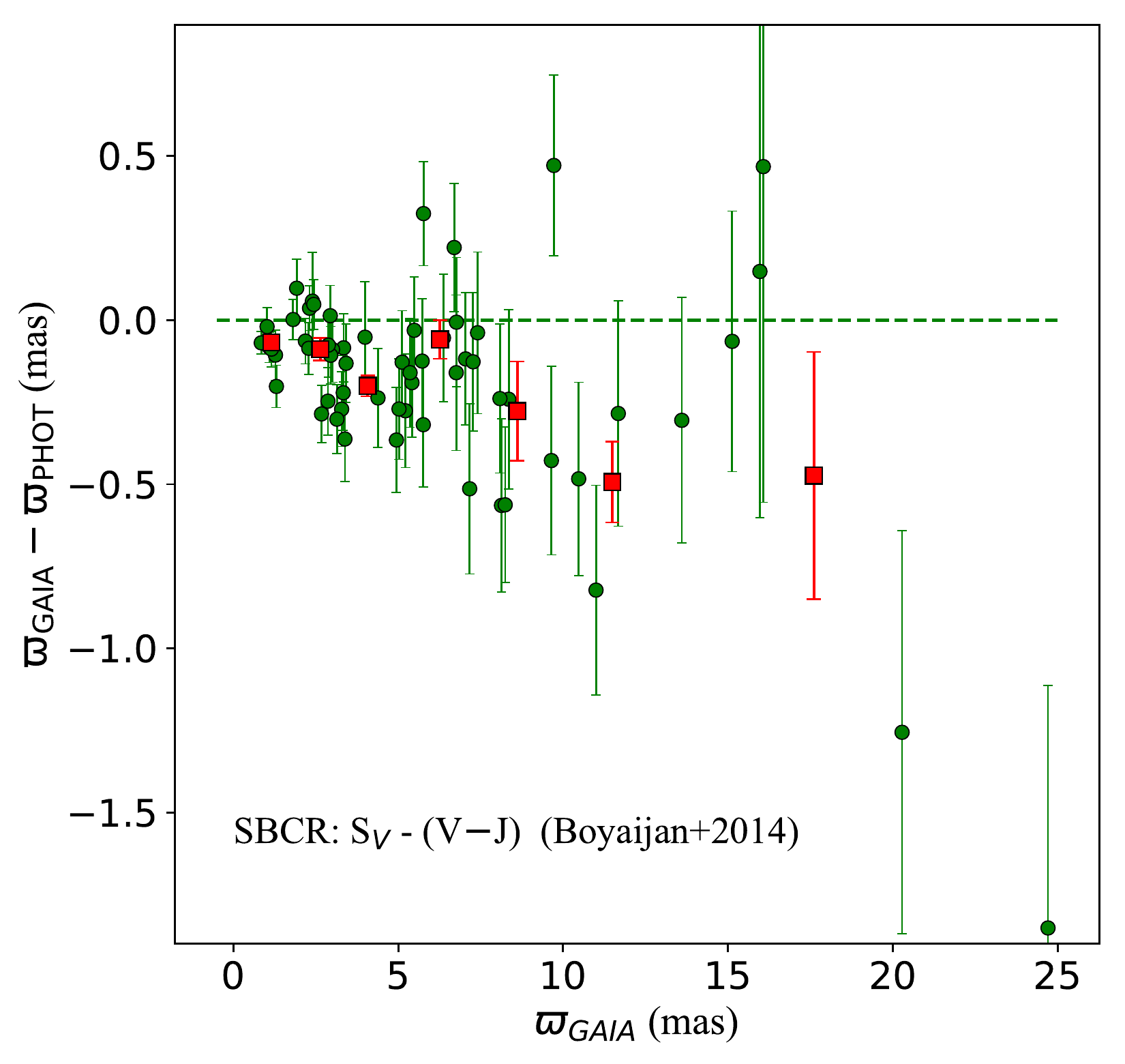}\\
\includegraphics[width=0.95\textwidth]{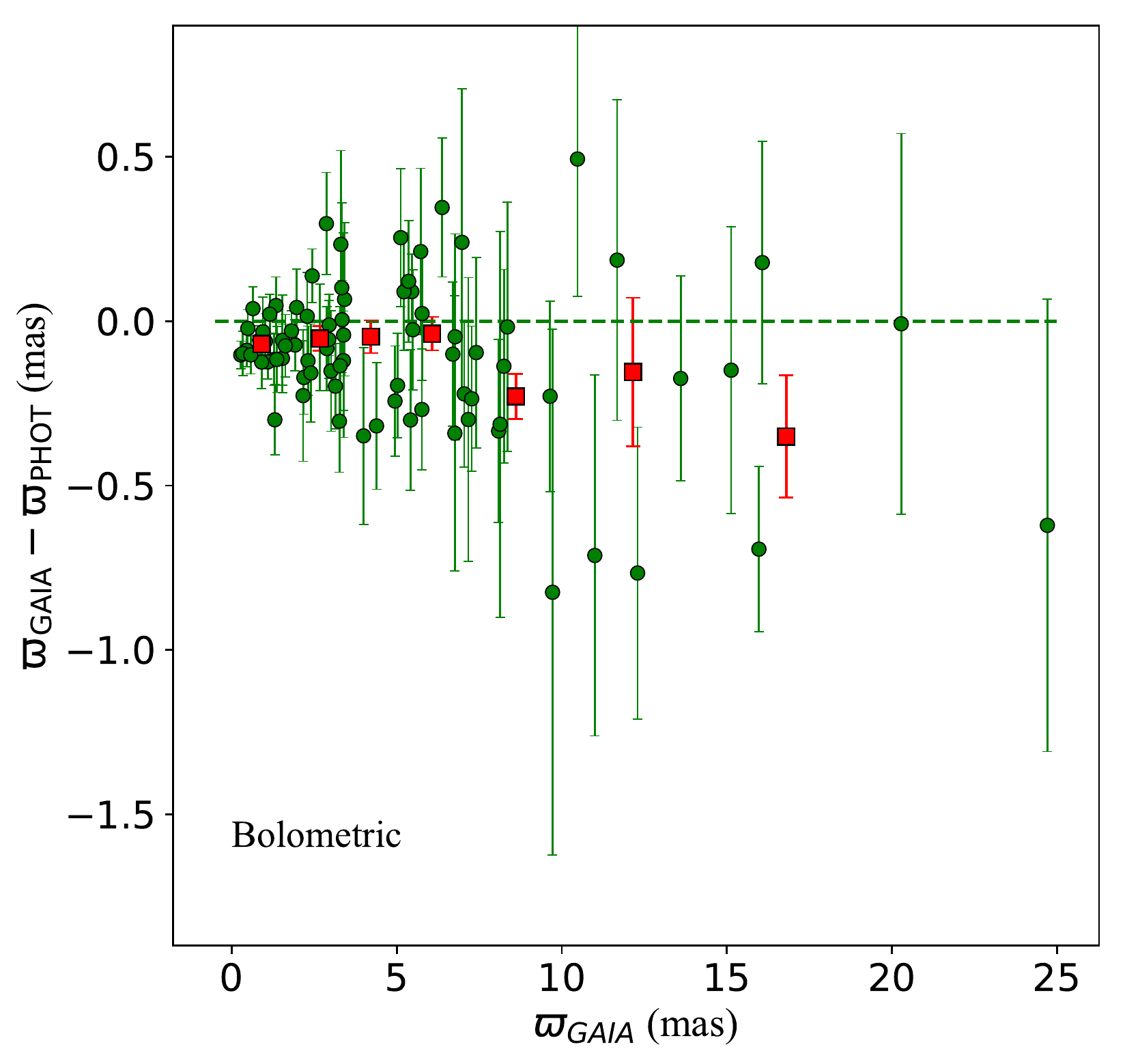} 
\end{minipage}
\caption{Individual parallax differences $d\varpi=\varpi_{Gaia}-\varpi_{Phot}$ of eclipsing binaries as a function of the {\it Gaia} parallaxes for all SBC relations used (green circles). The red squares are binned weighted means. The direct comparison of $ \varpi_{Gaia}$ and $\varpi_{Phot}$ for a few SBC relations is
given in Figure~\ref{fig:par}.  \label{fig:diff}}
\end{figure*}

\begin{figure*}
\centering
\begin{minipage}[t]{0.32\textwidth}
\includegraphics[width=0.95\textwidth]{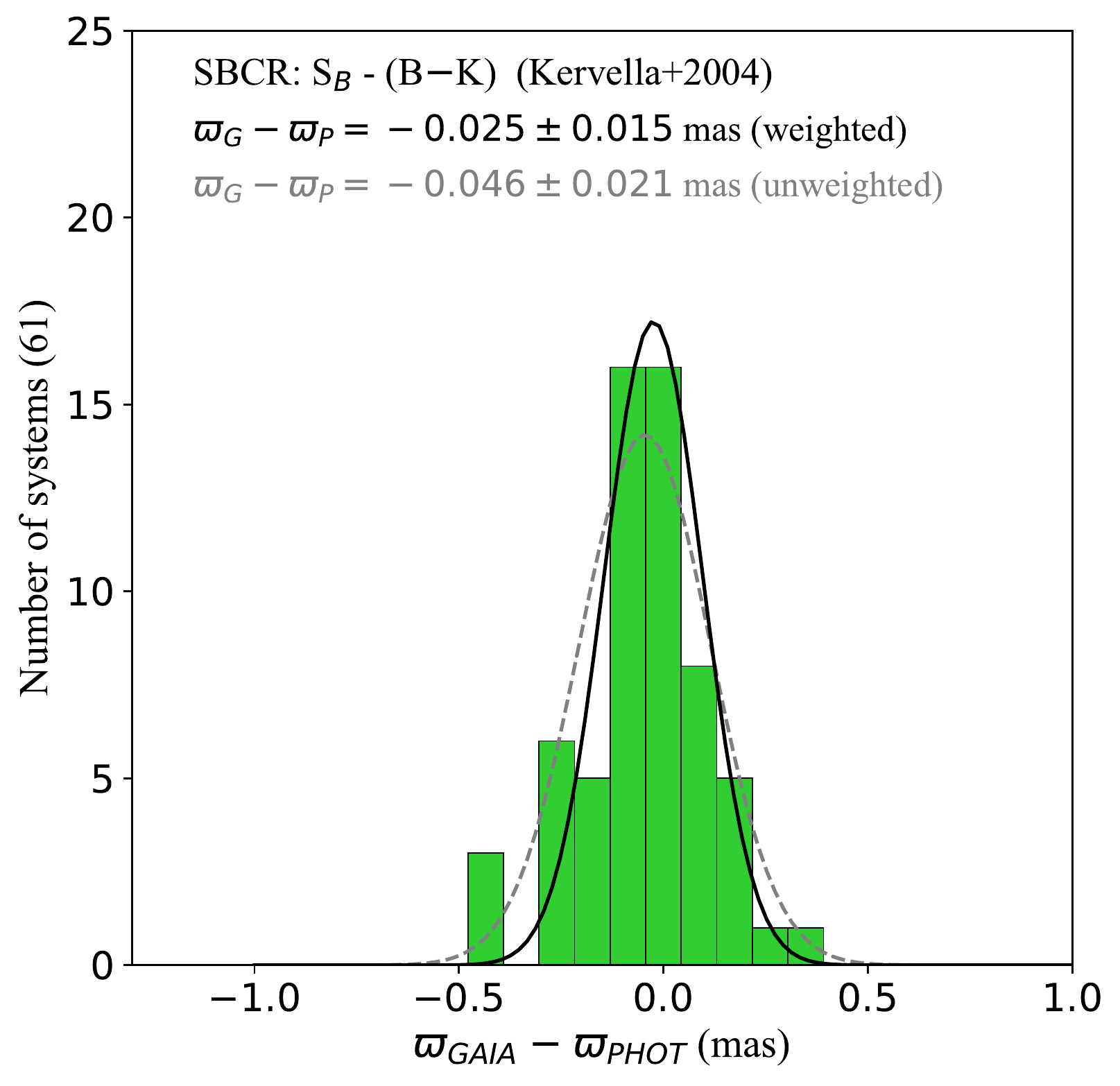} \\
\includegraphics[width=0.95\textwidth]{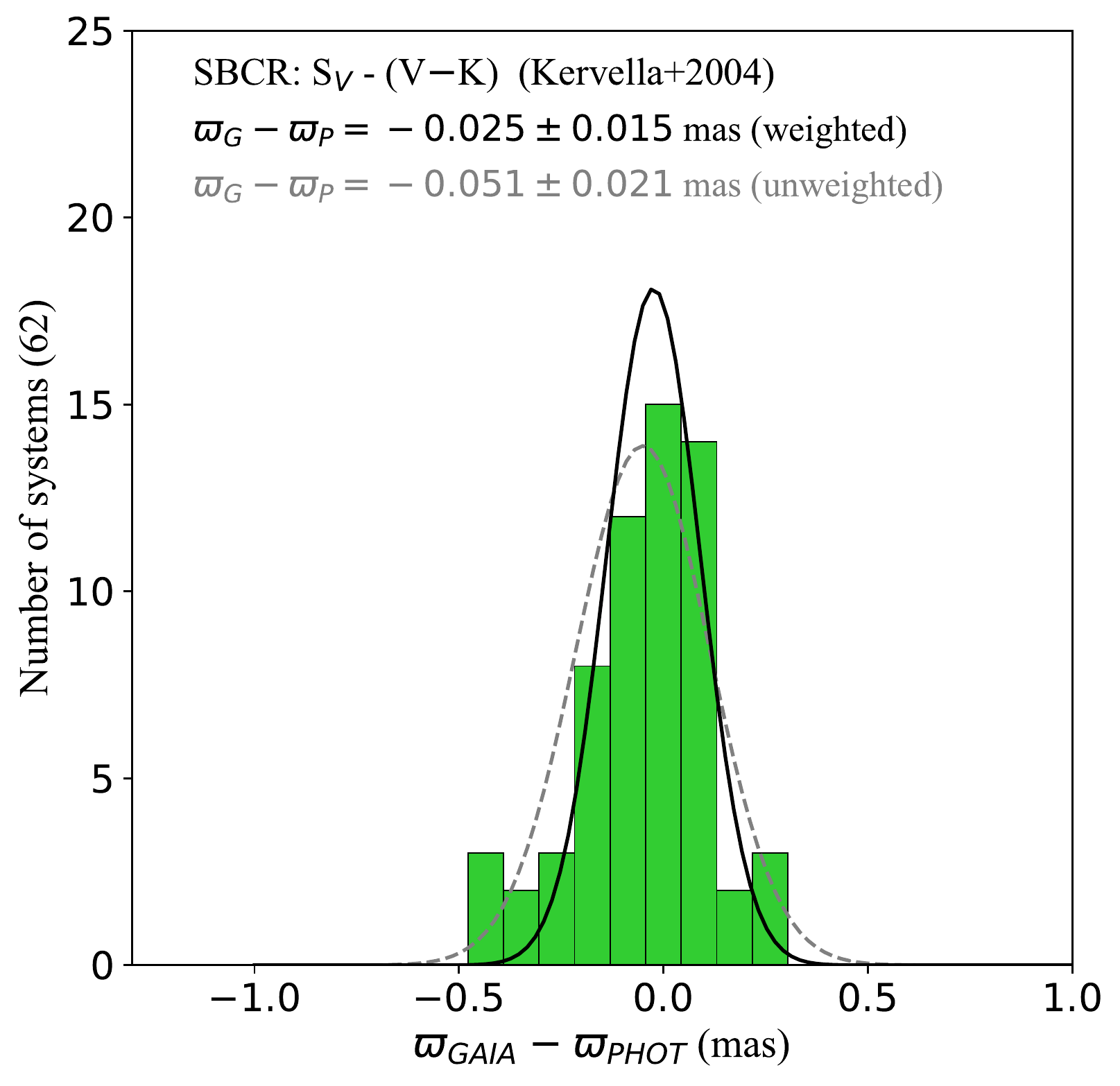} \\
\includegraphics[width=0.95\textwidth]{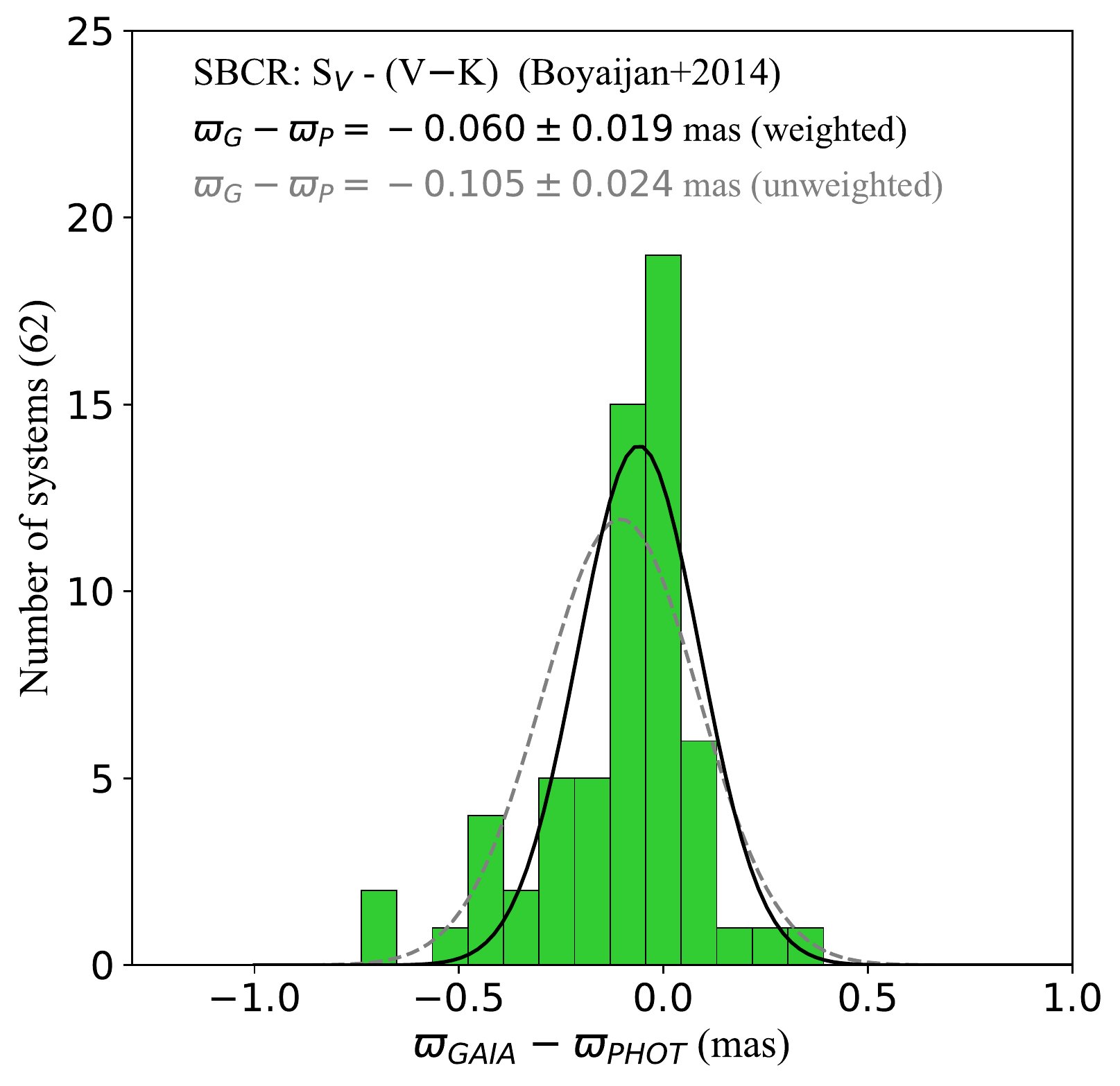} \\
\includegraphics[width=0.95\textwidth]{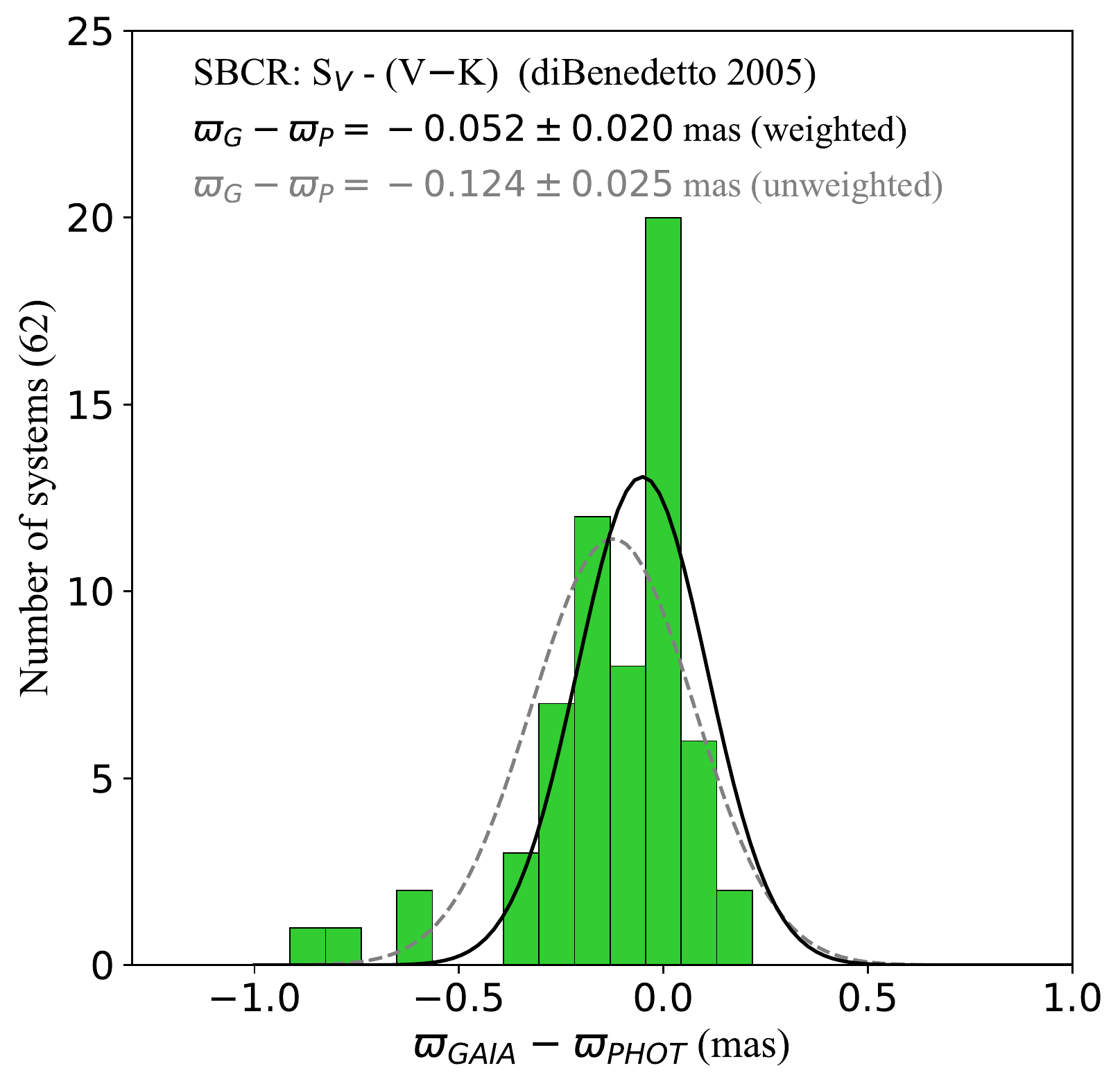}
\end{minipage}
\begin{minipage}[t]{0.32\textwidth}
\includegraphics[width=0.95\textwidth]{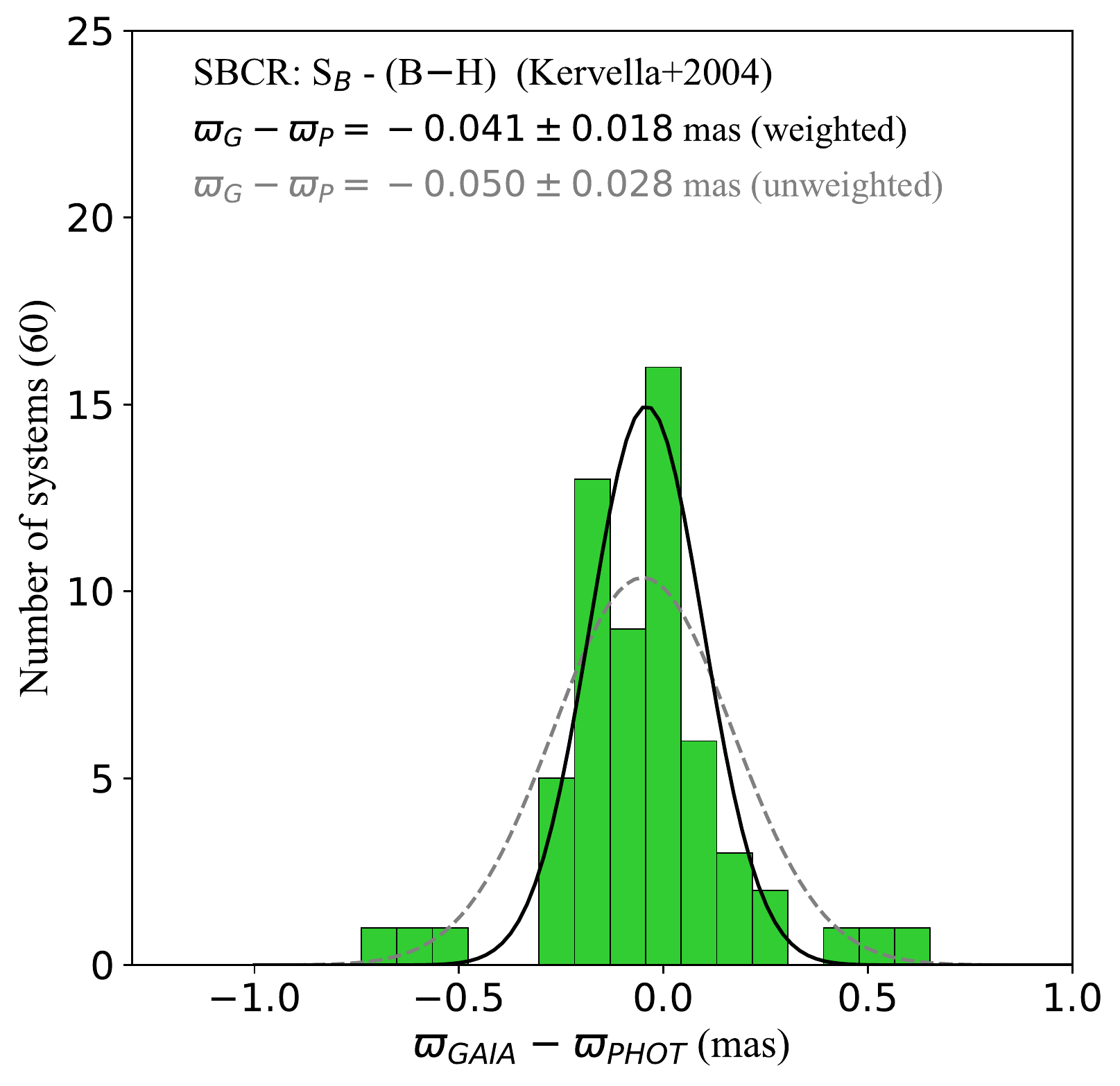}\\
\includegraphics[width=0.95\textwidth]{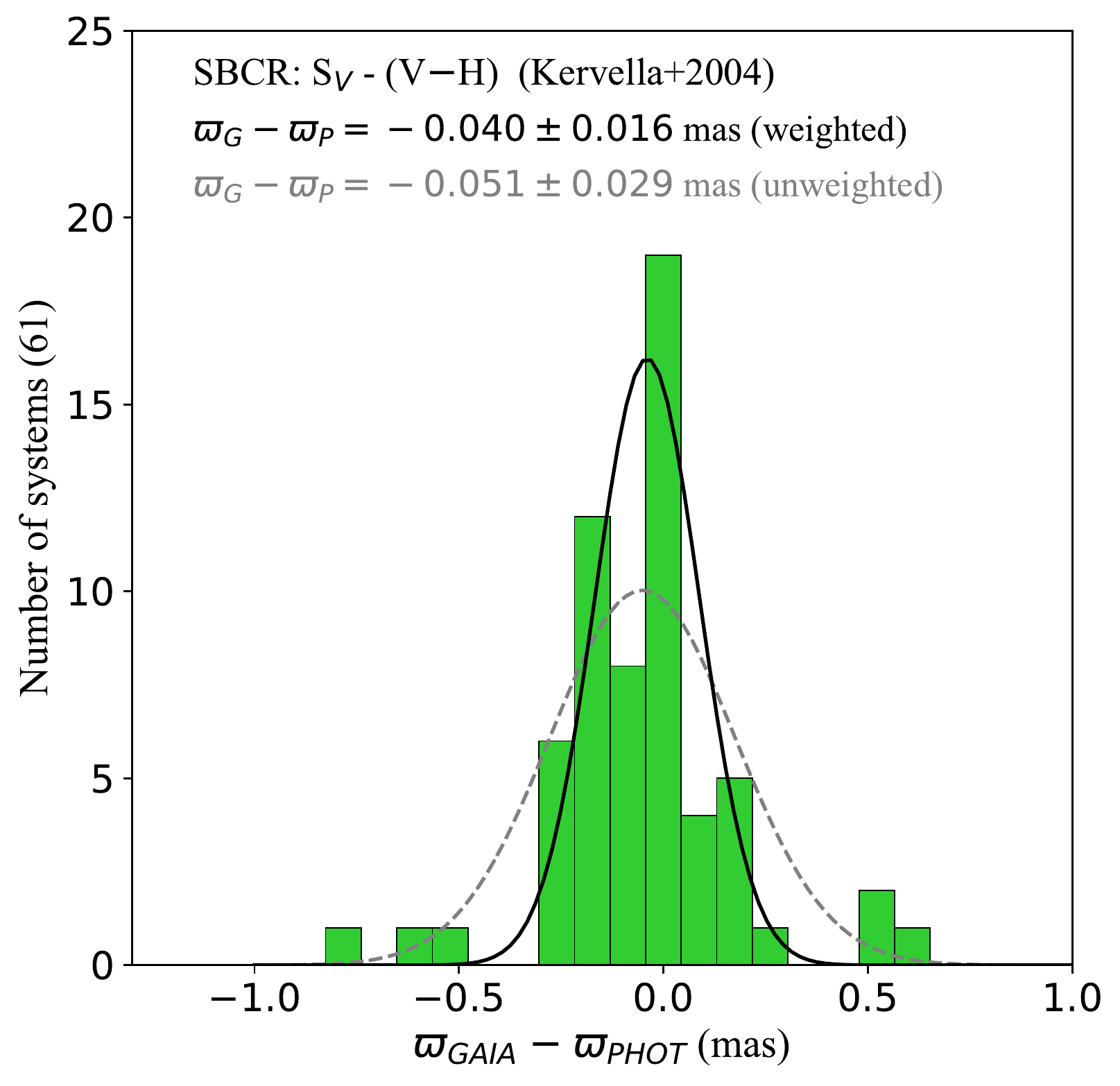}\\
\includegraphics[width=0.95\textwidth]{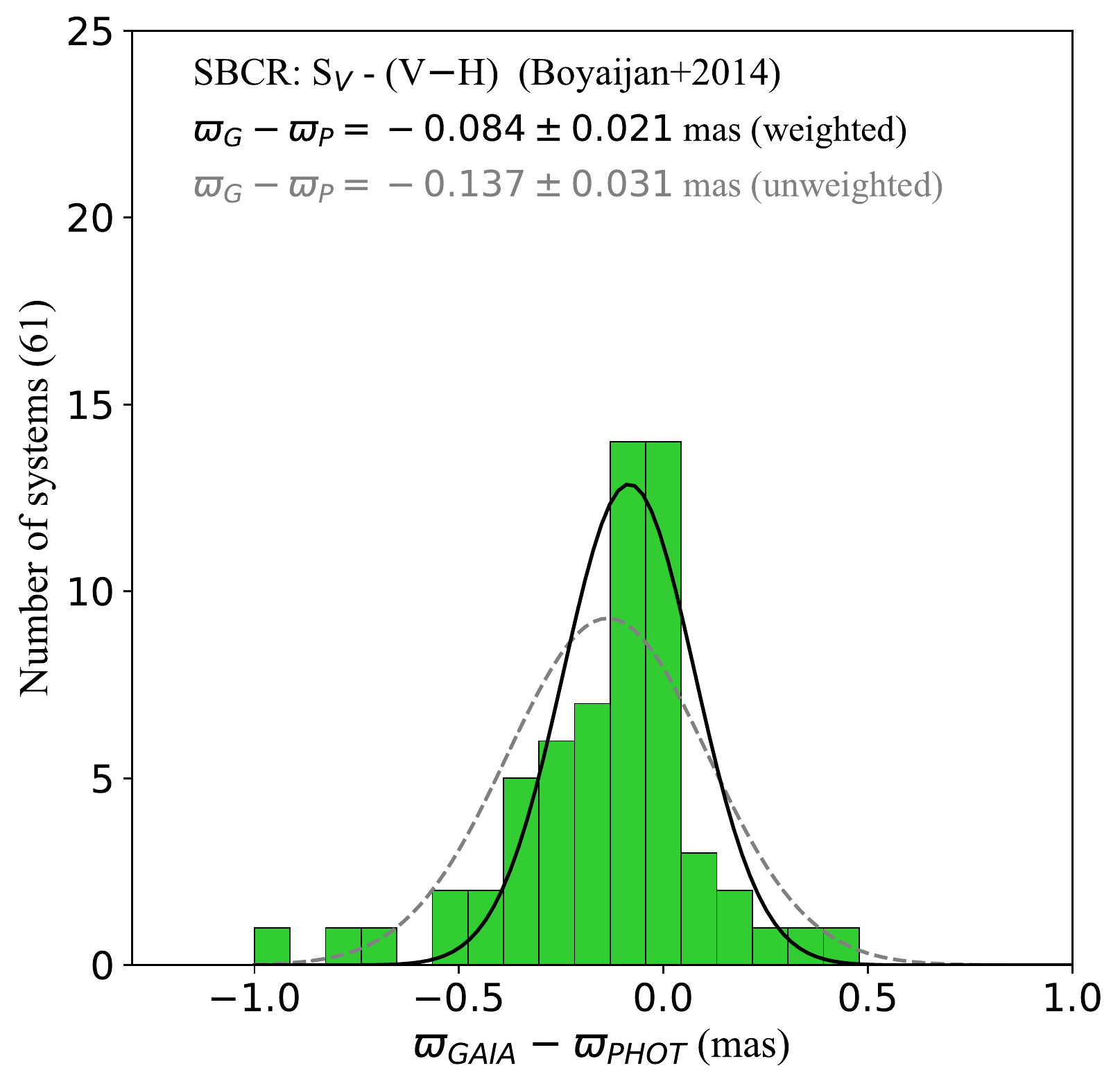}\\
\includegraphics[width=0.95\textwidth]{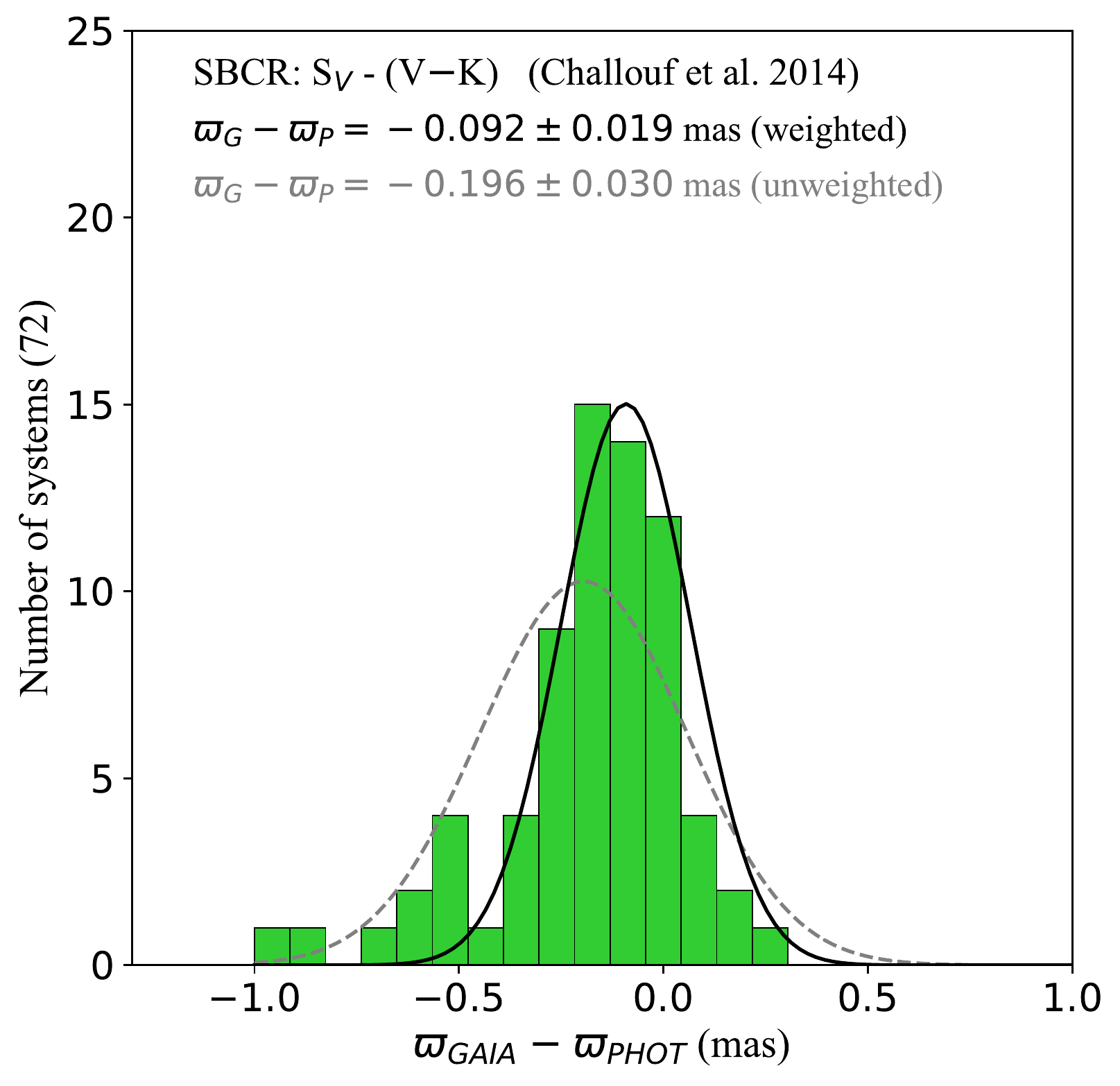}
\end{minipage}
\begin{minipage}[t]{0.32\textwidth}
\includegraphics[width=0.95\textwidth]{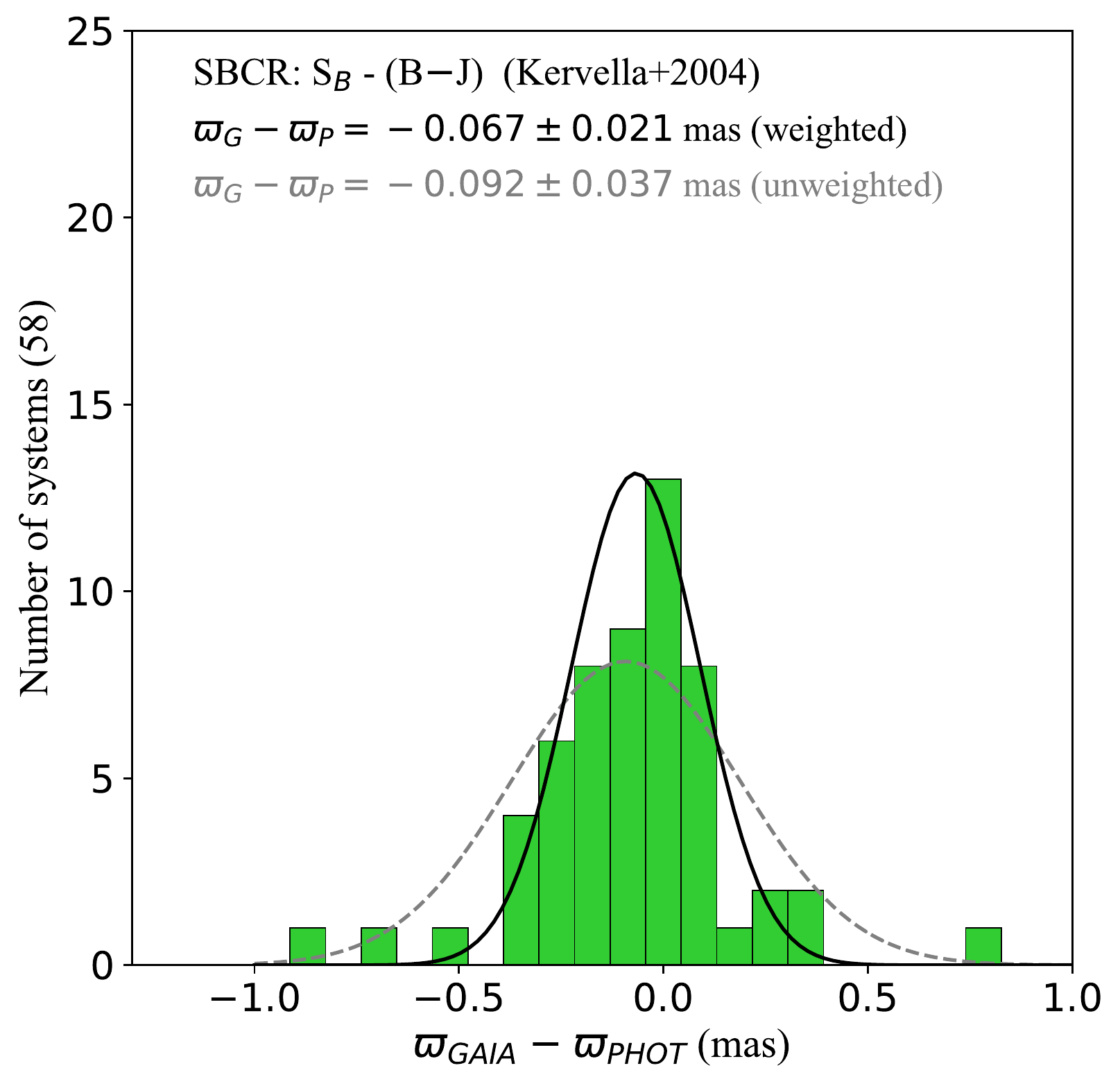}\\
\includegraphics[width=0.95\textwidth]{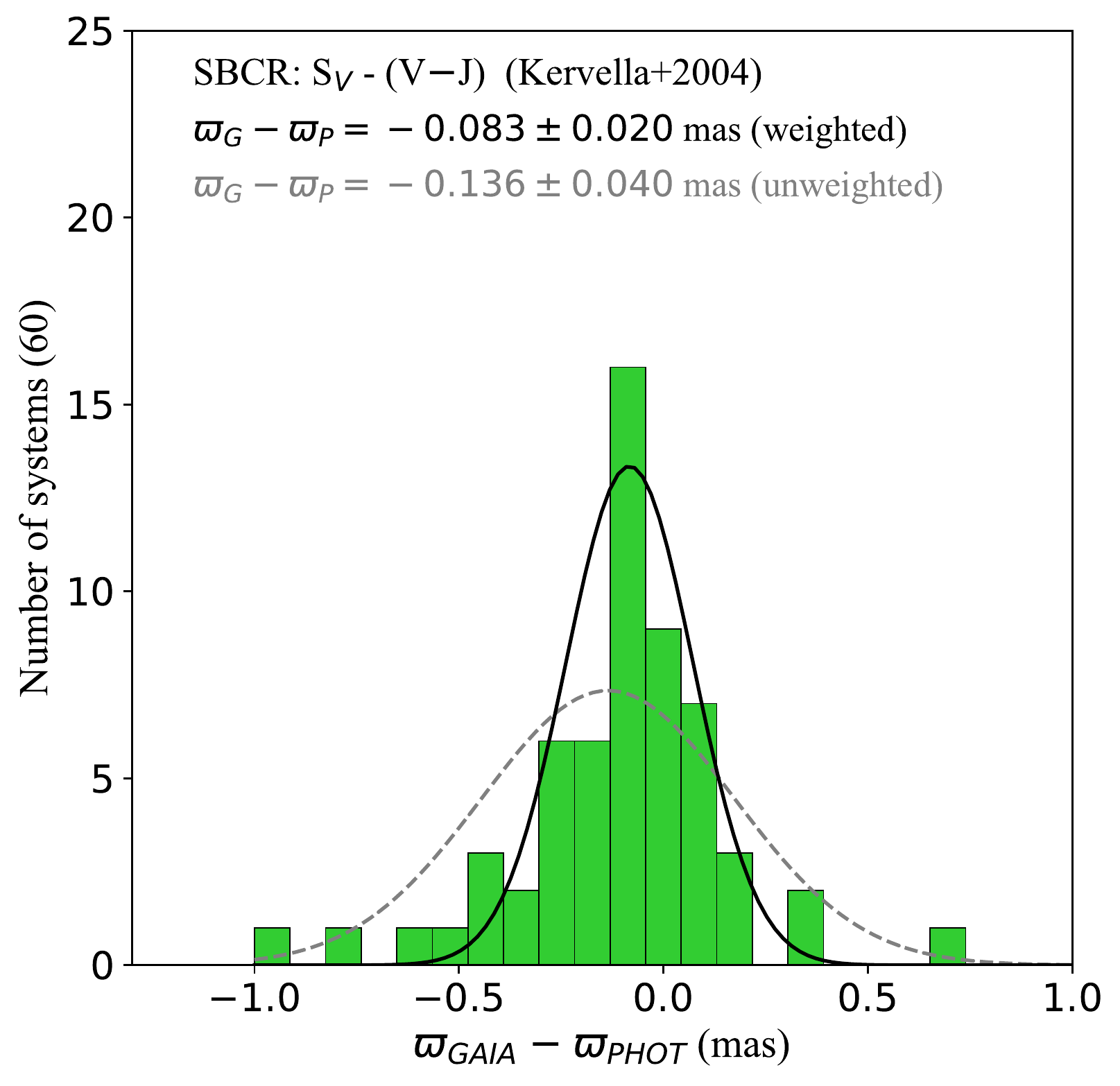}\\
\includegraphics[width=0.95\textwidth]{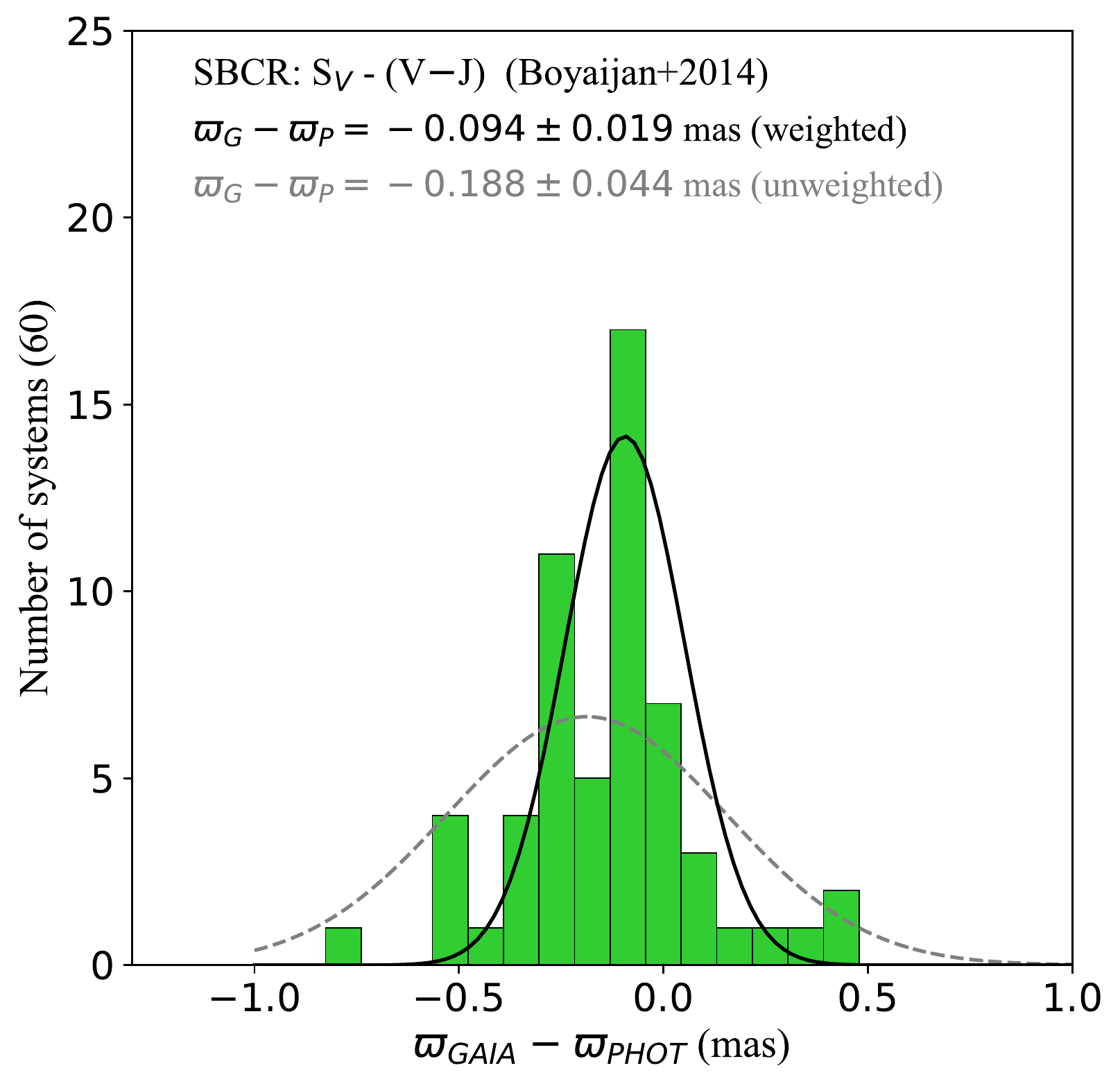}\\
\includegraphics[width=0.95\textwidth]{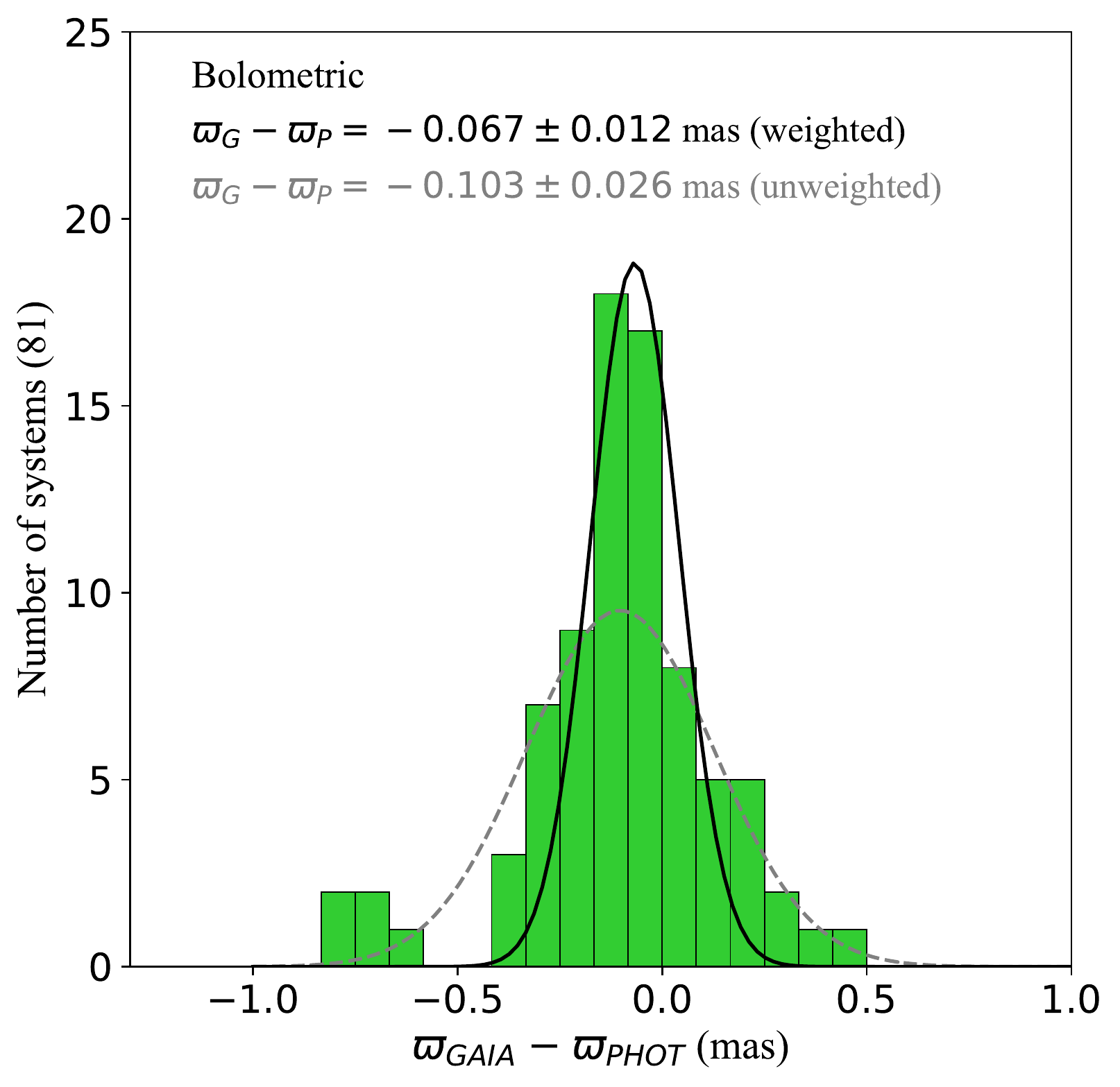} 
\end{minipage}
\caption{Histograms of the parallax differences $d\varpi=\varpi_{Gaia}-\varpi_{Phot}$ for all SBC relations used. The Gaussian distributions correspond to weighted (continuous line) and unweighted (broken line) mean values. They are given 
just for a reference. \label{fig:hist}}
\end{figure*}

\begin{figure*}
\centering
\begin{minipage}[t]{0.46\textwidth}
\includegraphics[width=0.95\textwidth]{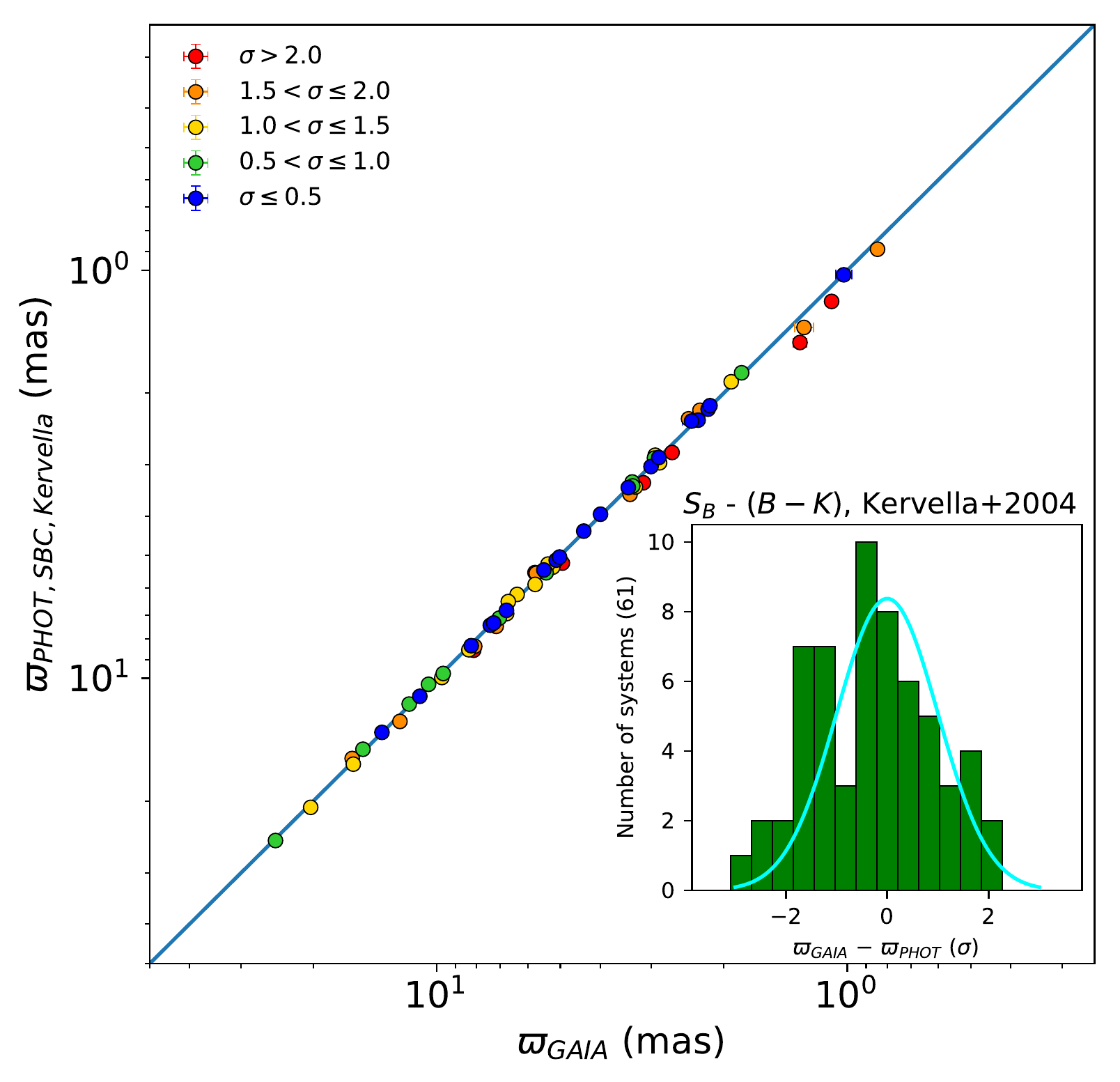} \\
\includegraphics[width=0.95\textwidth]{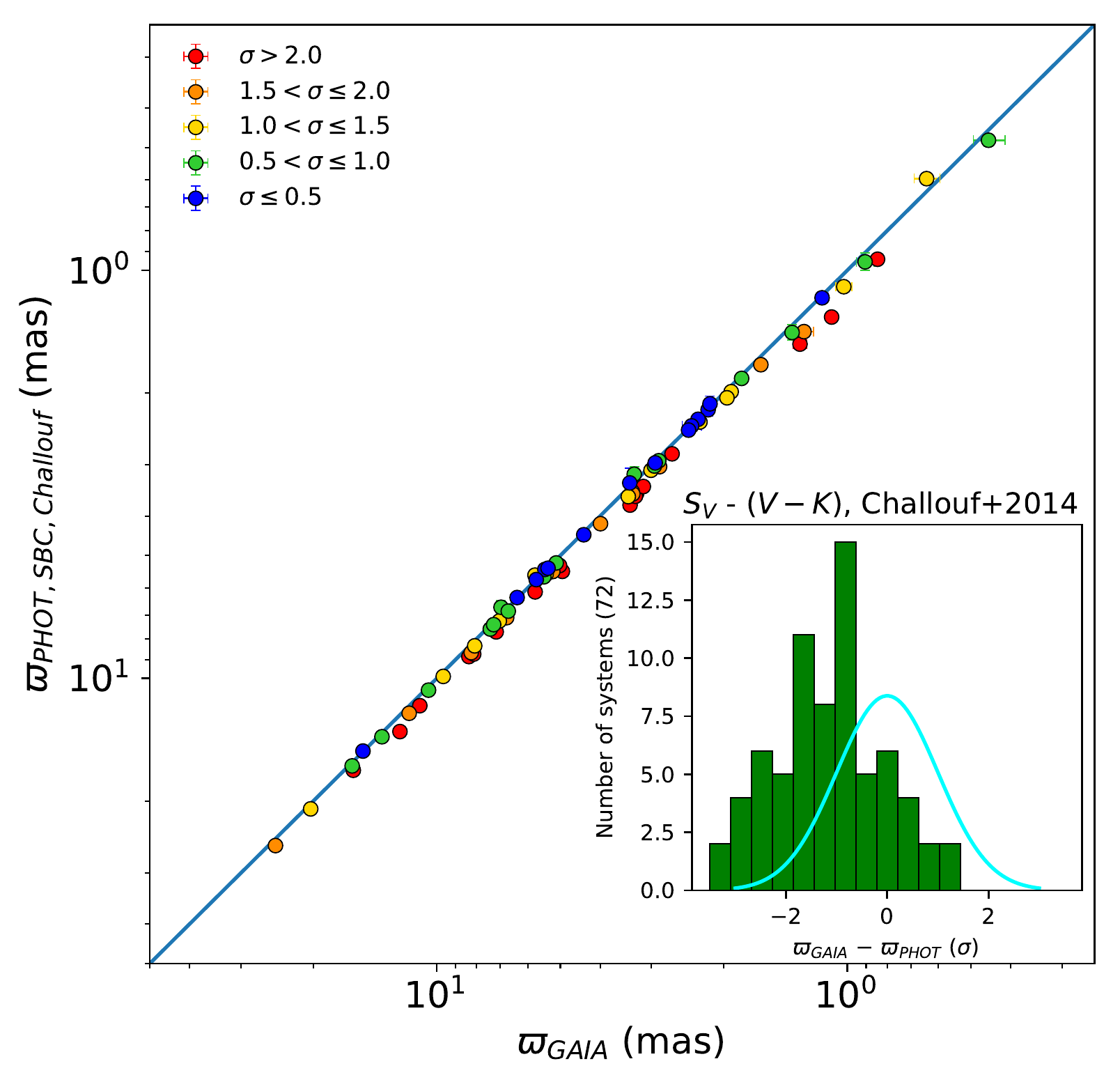} \\
\includegraphics[width=0.95\textwidth]{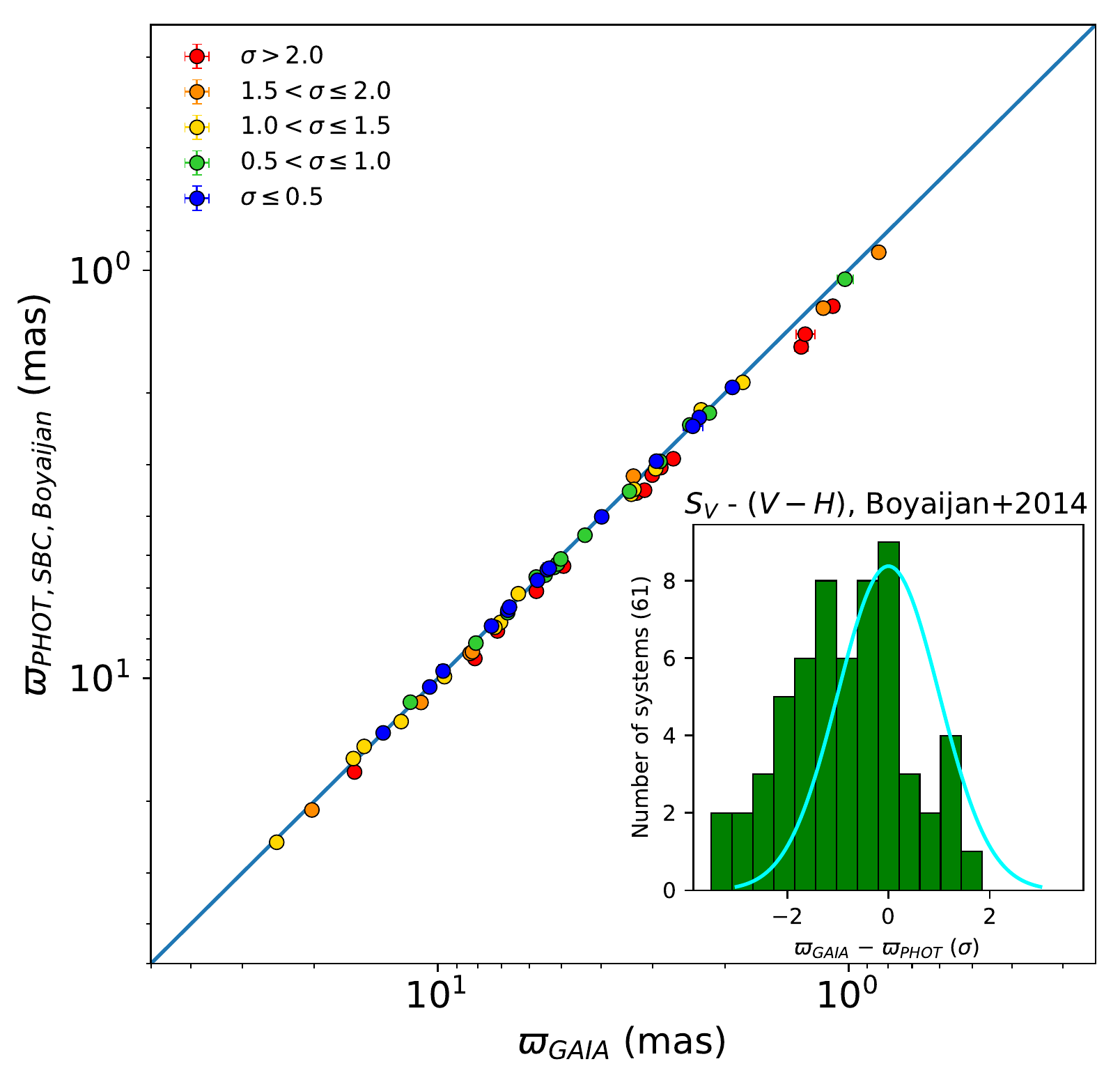}
\end{minipage}
\begin{minipage}[t]{0.46\textwidth}
\includegraphics[width=0.95\textwidth]{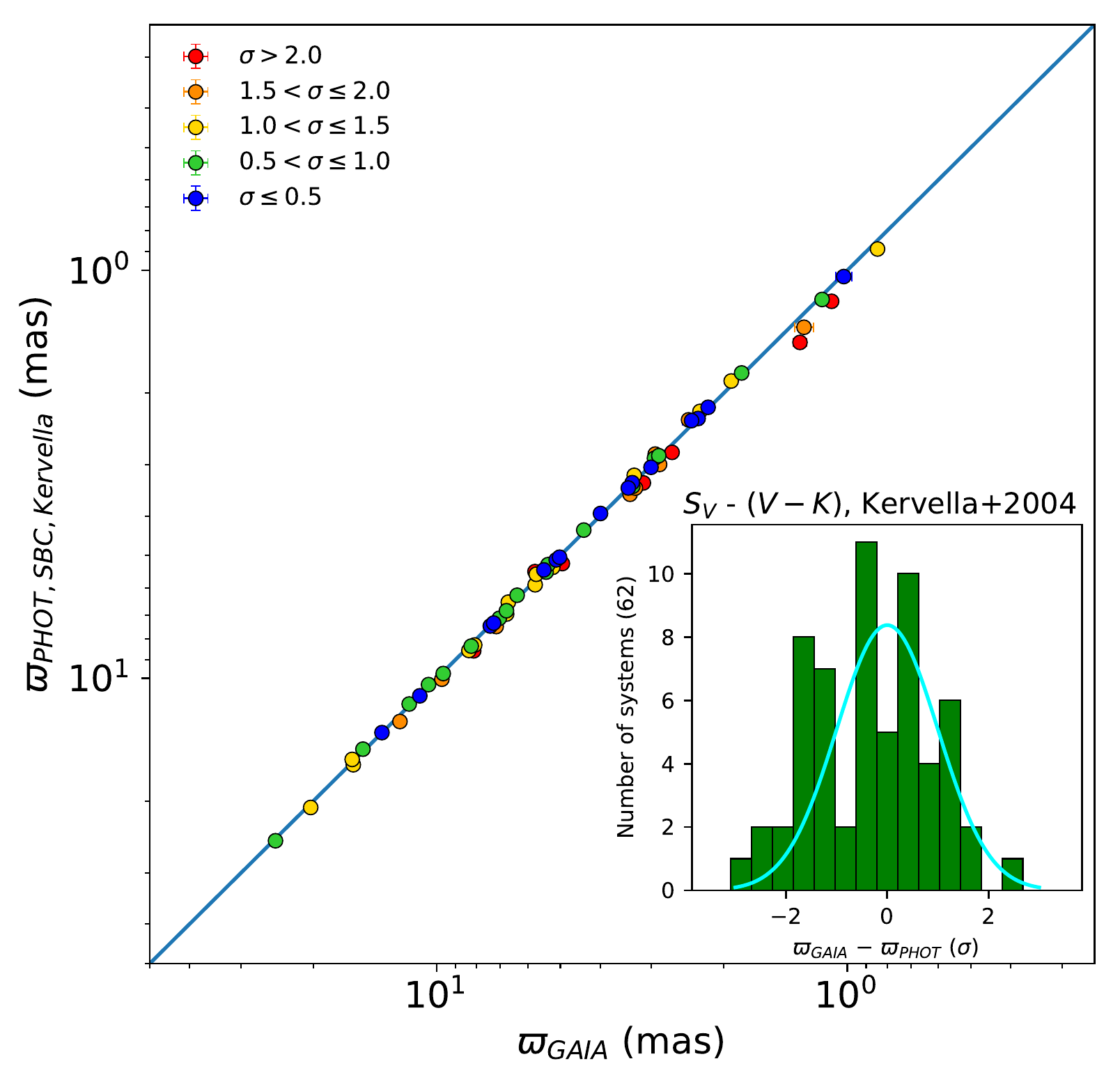} \\
\includegraphics[width=0.95\textwidth]{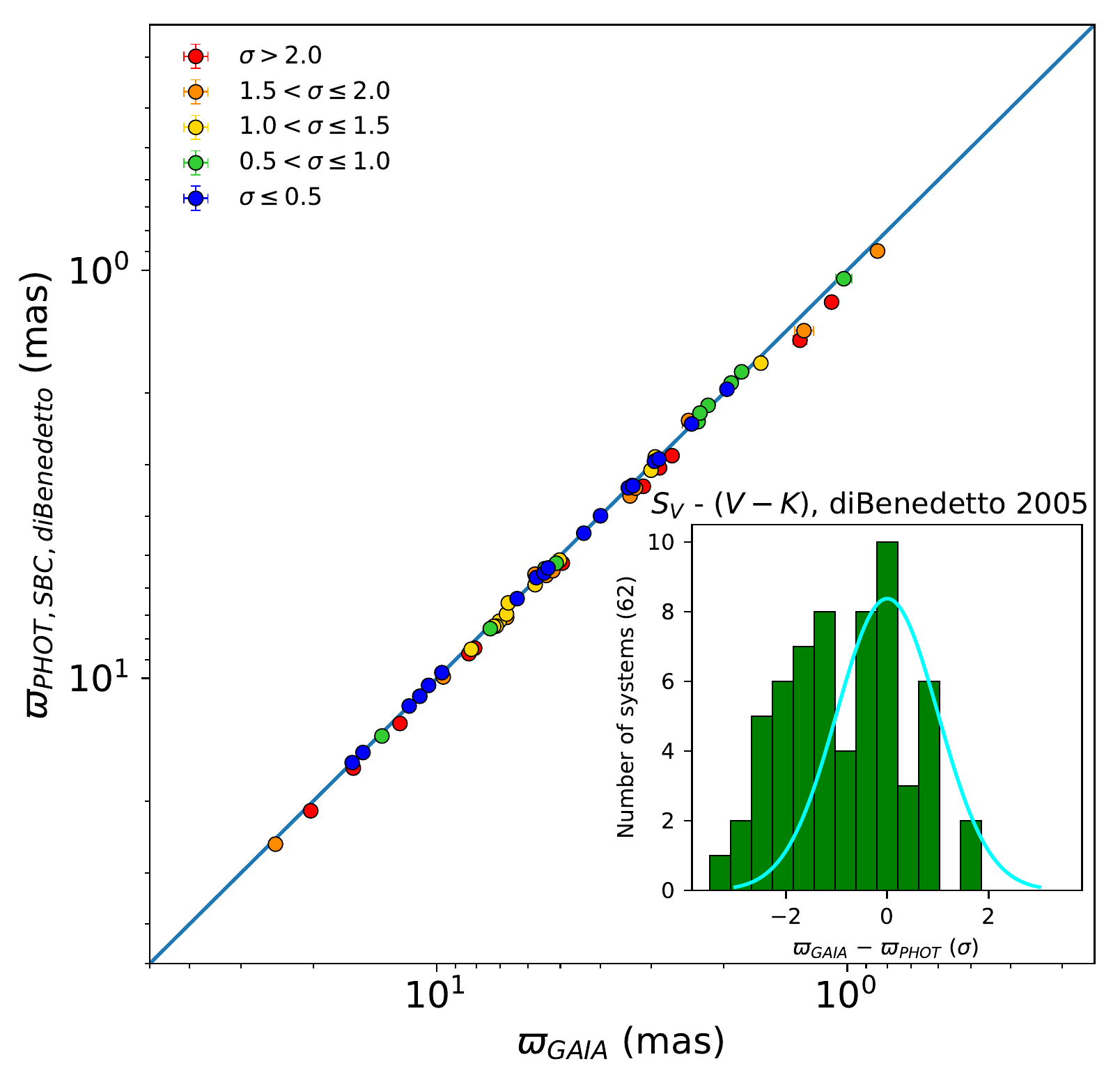} \\
\includegraphics[width=0.95\textwidth]{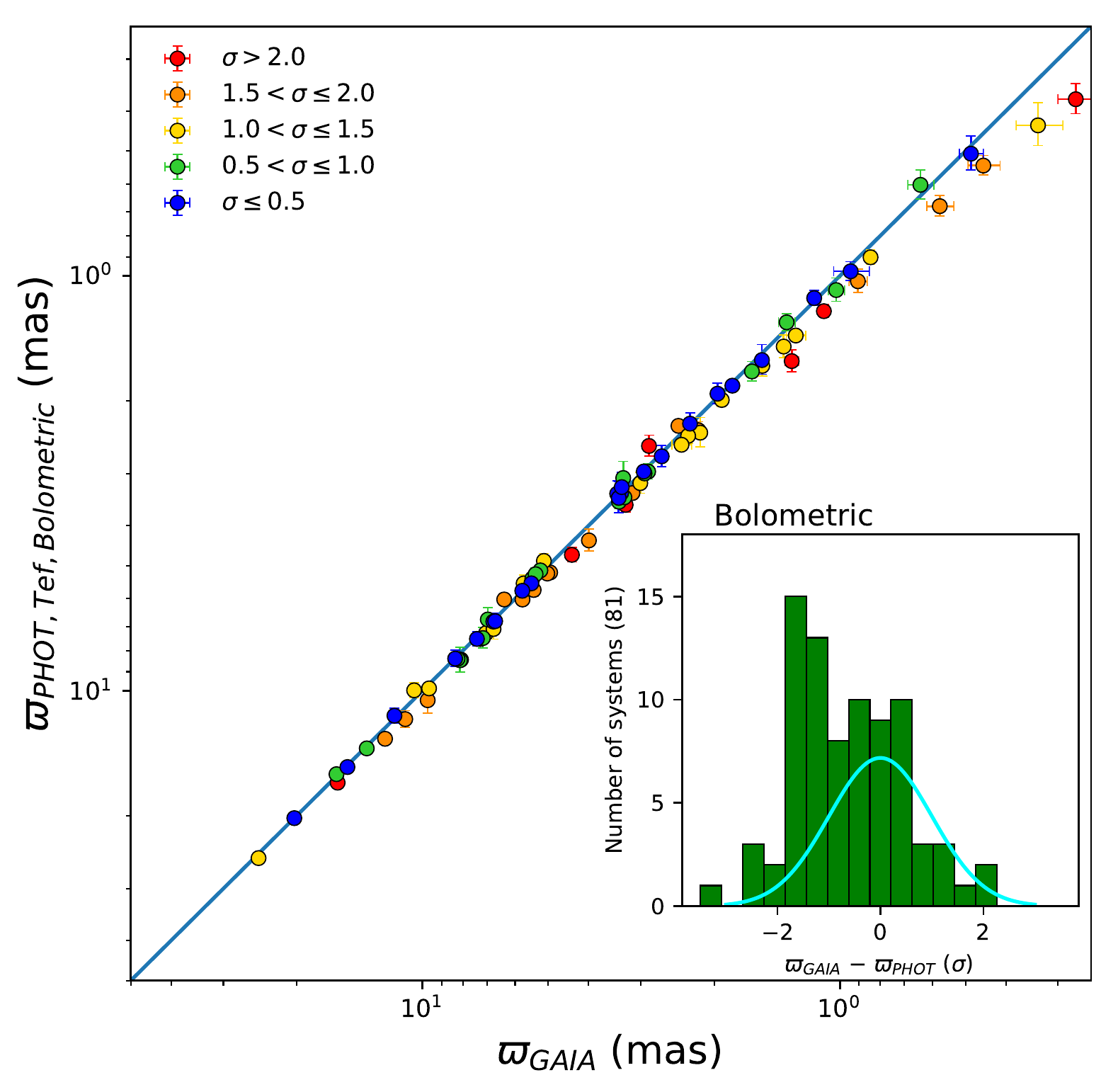} \\
\end{minipage}
\caption{Direct comparison of the {\it Gaia} and the photometric parallaxes for a few SBC relations. The errorbars in most cases are smaller than the size of the symbols. Insets show histograms like in Figure~\ref{fig:hist} but expressed
in terms of uncertainty of $d\varpi$. The Gaussians show expected distribution if no the offset would be present. \label{fig:par}}
\end{figure*}

\subsection{Calculation of the shift with respect to {\it Gaia} parallaxes}
We calculated individual parallax differences $ d\varpi_i = (\varpi_{Gaia} - \varpi_{Phot})_i$ for every $i$-th eclipsing binary and for each SBC relation used. 
For a given SBC relation we then determined the zero-point shift between the photometric and {\it Gaia} parallaxes as unweighted and weighted means of the differences.
The individual differences for all used SBC relations are presented in Fig.~\ref{fig:diff}. The errors of the differences were calculated taking into account systematic uncertainties of the SBC relations: 2\% for calibrations by \cite{ker04} and \cite{diB05}, 3\% for calibrations by \cite{boy14} and 3.5\% for calibration by \cite{cha14} -- their Eq.~13. Weighted means are dominated by more distant systems for which the error on $d\varpi$ is much smaller than for nearby eclipsing binaries. 
%\textcolor{red}
{This is a bit counterintuitive and it needs an explanation. Error on $d\varpi$ comes from two uncertainties: that of the {\it Gaia} parallax and that of the eclipsing binary photometric parallax. While the former for stars with $G < 14$ mag (all our sample) is almost independent of the distance and equal to 0.04 mas on average \citep{lind18}, the latter decreases with the distance because it is, in the majority of cases, dominated by a systematic uncertainty of the SBCR. Thus for a systematic error of 2\% on the photometric parallax the uncertainty is about 0.2 mas for a system lying 100 pc from the Sun and about 0.02 mas for a system lying 1 kpc away.}   
In Fig.~\ref{fig:hist} we present histograms of the differences for all the SBCR used, and also for parallaxes derived with the bolometric flux scaling method. The Gaussians correspond to a standard uncertainty distribution on weighted (continuous line) and unweighted means (dashed line). 

\section{Results}\label{results}
Tab.~\ref{tab:res} presents a summary of the zero-point $\varpi_{Gaia}-\varpi_{Phot}$ shifts for the different SBC relations, and in Fig.~\ref{fig:par} we show exemplary relations between trigonometric {\it Gaia} DR2 and photometric, eclipsing binary parallaxes.  The photometric parallaxes derived from bolometric fluxes correspond well with {\it Gaia} DR2 parallaxes, although the offset between both is clear. The amount of the offset of 
%\textcolor{red}
{$-0.067\pm0.012$ mas is marginally consistent with the zero-point shift reported by \cite{sta18}. Our sample agrees in about 80\% with a sample used by \cite{sta18}, we also adopted the same solar constants; however, we used a different method to derive $L_{bol}$ which may account for the difference with their result.} We point out that the accuracy of the derived offset is limited by the accuracy of the zero-points of the different methods used to determine the effective temperatures 
%\textcolor{red}
{and the choice of solar constants.} However, the problem of homogenization of the temperature determination for all the sample and the proper calibration of the temperature zero-point scale is beyond the scope of this paper. We therefore prefer to establish the {\it Gaia} zero-point shift solely based on the empirical SBC relations, which allows for a homogenous treatment of the sample (and its subsamples).  

As pointed out already by \cite{sta18} there is almost perfect agreement between the photometric and {\it Gaia} DR2 parallaxes. However, some of the SBC relations result in a larger offset and larger differences \citep[e.g.][]{cha14} -- see Fig.~\ref{fig:diff} and Fig.~\ref{fig:hist}. The possible reasons for this are discussed in the next section.   

The SBC relations, which result in the smallest differences and best agreement with {\it Gaia} DR2 parallaxes are those based on $(B\!-\!K)$ and $(V\!-\!K)$ colors and calibrated by \cite{ker04}. They are also the relations that result in the smallest internal dispersions of the differences. Calibrations based on these two colors are the least reddening dependent because the reddening line on the surface brightness $S$ - color diagram is almost parallel to the relations themselves. Moreover, the calibrations by \cite{ker04} were done for main-sequence dwarfs and subgiants, which constitute the overwhelming majority of the eclipsing binary component stars in our sample. The resulting zero-point shift based on the
$(B-K)$ and $(V-K)$ relations is $-0.025\pm0.011$ mas. Relations by \cite{ker04} calibrated on the other optical-infrared colors are more reddening dependent and also show significantly larger scatter of the differences. 

Another SBC relation that we used extensively for the determination of the distances to the Magellanic Clouds \citep[][e.g.]{pie13,gra14} was calibrated by \cite{diB05}. This relation is calibrated on a mixture of giant and dwarf stars. The resulting zero-point shift is $d\varpi=-0.052\pm0.020$ mas and is within $1.3\sigma$ consistent with the $d\varpi$ derived from the Kervella et al. SBC relations. Combining these two determinations we obtain a value of $d\varpi=-0.031\pm0.011$.

\begin{deluxetable*}{@{}lcccccc@{$\,\pm\,$}cc@{$\,\pm\,$}c@{}}
\tabletypesize{\scriptsize}
\tablecaption{The zero point shifts $\varpi_{Gaia}-\varpi_{Phot}$ determined with eclipsing binaries for the different SBCRs \label{tab:res}}
\tablewidth{0pt}
\tablehead{
\colhead{SBC} & \colhead{Band} & \colhead{Color} & \colhead{Range} & \colhead{$\log{g}$} & \colhead{Number of} & \multicolumn{4}{c}{($\varpi_{Gaia}-\varpi_{Phot}$) (mas)}\\
\colhead{relation} & \colhead{} &\colhead{} & \colhead{of color} & \colhead{(dex)} & \colhead{systems} & \multicolumn{2}{c}{Unweighted}  & \multicolumn{2}{c}{Weighted}}
\startdata
Kervella et al. 2004 &$B$ &$B\!-\!K$ & $-$0.25 - 2.36 & $>\!3.5$ & 61 & $-0.046$ &0.021 & $-0.025$ & 0.015 \\
Kervella et al. 2004 &$B$ &$B\!-\!H$ & $-$0.20 - 2.28  & $>\!3.5$ & 60 & $-0.050$ &0.028 & $-0.041$ & 0.018 \\
Kervella et al. 2004 & $B$&$B\!-\!J$ & $-$0.17 - 2.08  & $>\!3.5$ & 58 & $-0.092$ &0.037 & $-0.067$ & 0.021 \\
Kervella et al. 2004 &$V$ &$V\!-\!K$ & $-$0.20 - 1.75  & $>\!3.5$ & 62 & $-0.051$ &0.021 & $-0.025$ & 0.015\\
Kervella et al. 2004 &$V$ &$V\!-\!H$ & $-$0.15 - 1.66  & $>\!3.5$ & 61 & $-0.051$ &0.029 & $-0.040$ & 0.016\\
Kervella et al. 2004 &$V$ &$V\!-\!J$ & $-$0.12 - 1.47  & $>\!3.5$ & 60 & $-0.136$ &0.040 & $-0.083$ & 0.020\\
\\
di Benedetto 2005 &$V$ &$V\!-\!K$ & $-$0.10 - 4.93  & $>\!2.0$ & 62 & $-0.124$ &0.025 & $-0.052$ & 0.020\\
\\
Boyajian et al. 2014 &$V$ &$V\!-\!K$ & $-$0.15 - 1.75  & $>\!3.5$ & 62 & $-0.105$ &0.024 & $-0.060$ & 0.019\\
Boyajian et al. 2014 &$V$ &$V\!-\!H$ & $-$0.13 - 1.66  & $>\!3.5$ & 61 & $-0.137$ &0.031 & $-0.084$ & 0.021\\
Boyajian et al. 2014 & $V$&$V\!-\!J$ & $-$0.12 - 1.47  & $>\!3.5$ & 60 & $-0.188$ &0.044 & $-0.094$ & 0.019\\
\\
Challouf et al. 2014 &$V$&$V\!-\!K$ & $-$0.60 - 4.93 & $>\!2.0$ & 72 & $-0.196$ &0.030 & $-0.092$ & 0.019\\
\\
Bolometric  & - & - & - & - & 81 & $-0.103$ & 0.026 & $-0.067$ & 0.012  
\enddata
\tablecomments{A blue edge of the color range is constrained by the SBC relation color range validity (with an exception of relation by \cite{cha14}) and a red range is contrained by most red systems in the sample.}
\end{deluxetable*}

\section{Discussion}
Using the most reliable SBC relations available in the literature, we derived a zero-point shift for {\it Gaia} DR2 parallaxes of $d\varpi=-0.031\pm0.011$ mas. However, there are significant differences of the zero-point shifts resulting from applying 1) different SBC calibrations, and 2) different colors. 

The reasons for the existence of the SBC calibration dependent zero-point shifts are somewhat unclear. For sure a choice of different interferometric stellar angular diameter sets will result in different calibrations. Most notably a calibration derived from a mixture of all kinds of stars \citep[e.g.][]{cha14} may lead to a systematic difference with a calibration based on a mixture of dwarf and subgiant stars only \citep[e.g.][]{ker04}. However, more worrisome is a systematic difference of the zero-point shifts derived from the same colors but from two different calibrations \citep{ker04,boy14} that are based on angular diameters of dwarf and subgiants stars and calibrated on the Johnson system. 
%\textcolor{red}
{We searched relevant papers to find the differences. \cite{ker04} use a small but well characterized sample of stars, while \cite{boy14} used a much larger sample but containing multiple and variable stars, and also utilizing saturated 2MASS photometry for some bright stars. Because of this we tend to prefer the former calibrations over the latter.} The systematic differences between the calibrations (see Tab.~\ref{tab:res}) vary from $-0.011$ mas to $-0.044$ mas with a weighted average of $-0.031$ mas. 

\begin{figure*}
\centering
\mbox{\includegraphics[width=0.4\textwidth]{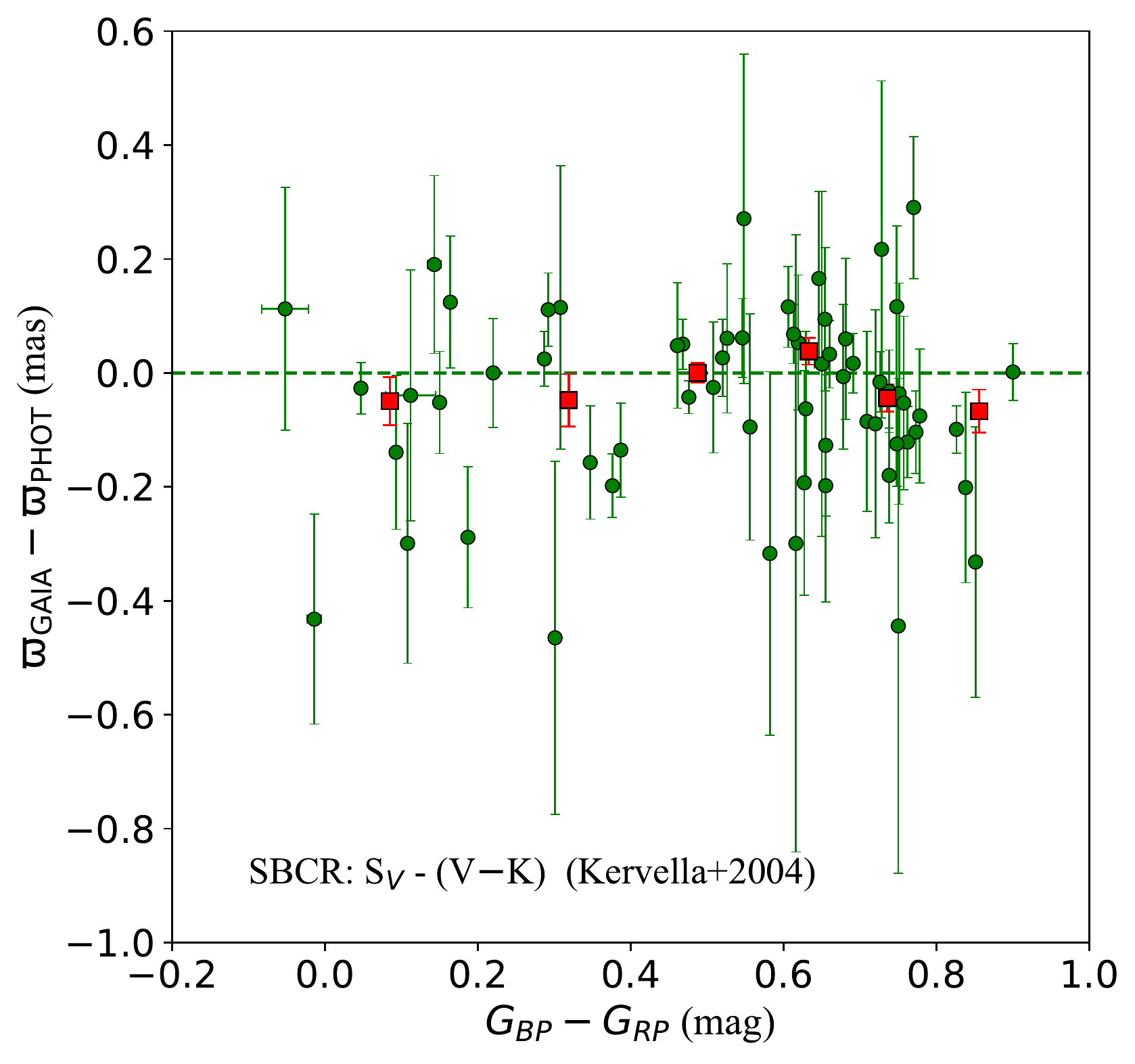}} 
\mbox{\includegraphics[width=0.4\textwidth]{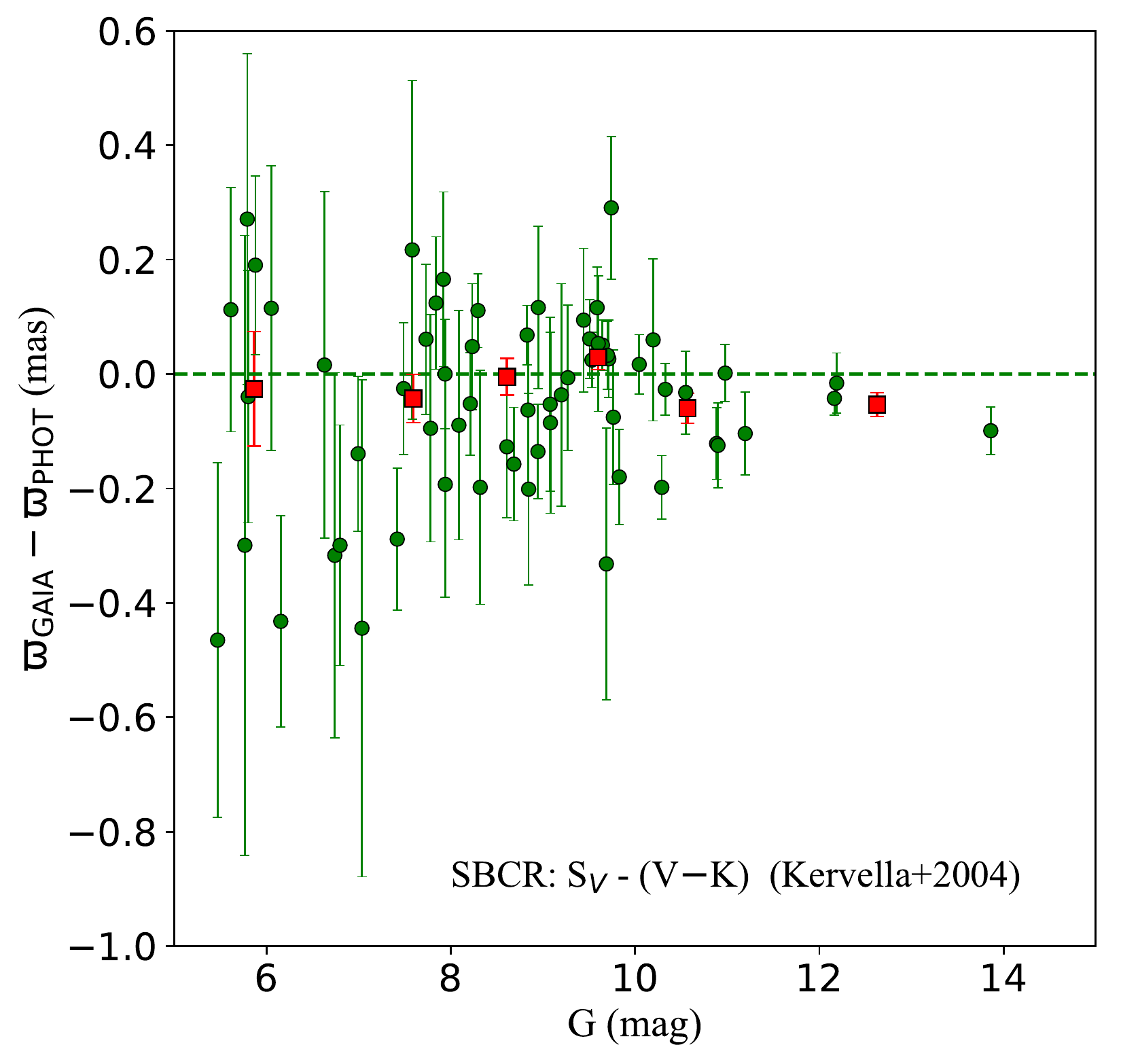}} 
\caption{Color and magnitude dependency of the parallax differences for the SBC relation by \cite{ker04}. The meanings of the symbols are the same as in Fig.~\ref{fig:diff}}
\label{fig:func}
\end{figure*}

Regarding 2) there is a clear trend that relations calibrated on shorter-wavelength colors (and therefore being more reddening dependent) give systematically larger zero-point shifts and larger scatter of the differences, with the largest shifts coming from SBC relation calibrations based on the $(V\!-\!J)$ color. We tried to remove the shifts by a rescaling of all individual reddening estimates according to $E(B\!-\!V)_{\rm new}=C\cdot E(B\!-\!V)$, where $C>1$ is a scaling factor. Unless $C$ is larger then about 4 there is no possibility to obtain an agreement between zero-point shifts derived from different colors. But such a large value of the scaling factor would lead to an unrealistically large extinction for most of the targets in our sample, and would be strongly at odds with independent reddening determinations by \cite{sta16}. Another, but less likely, possibility is that the transformation equations from Sec.~\ref{sec:NIR} contain some systematic error leading to color dependent zero-point shifts. And at last the SBC relations calibrated on colors containing $J$ and $H$ may be problematic by themselves because of a proper calibration of the ground-based photometry in these filters is difficult (strong and variable atmospheric extinction) and because of stronger sensibility to the interstellar reddening.

We investigated how the global shift depends on a distance. We binned up individual $d\varpi_i$ in several bins -- see Fig.~\ref{fig:diff}; however, no clear systematic trend can be noticed. We also investigated a possible {\it Gaia} color $G_{BP}-G_{RP}$ and $G$ magnitude dependency. We plotted $d\varpi$ differences for a one specific SBC relation -- Fig.~\ref{fig:func}. As previously we do not detect any systematic trend. 
%\textcolor{red}
{We also investigated the correlation of $d\varpi$ with sky position. We calculated separately shifts for equatorial and ecliptic hemispheres. The differences between hemispheres are in  both cases within 1$\sigma$ uncertainty and thus not statistically significant.}

The comparison with previous determinations of $d\varpi$ shows excellent agreement with the {\it Gaia} Team results \citep{lind18,are18}. In fact we confirm here the zero-point shift of $d\varpi=-0.029$ mas reported by the {\it Gaia} Team. However, comparisons with other independent determinations show some discrepancies. Although the $d\varpi$ value reported by \cite{ries18} agrees with our value at the $1\sigma$ level, the zero-point shifts reported by \cite{sta18} and \cite{zinn18} are somewhat discrepant.  The problem with the shift reported by \cite{zinn18} is its unrealistically small systematic uncertainty, which ignores systematics of asteroseismic relations themselves and the zero-point uncertainty of the temperature scale used by the authors. Because of this a comparison of the shift by \cite{zinn18} with other reported shifts is vague. 
 
\begin{figure}
\mbox{\includegraphics[width=0.44\textwidth]{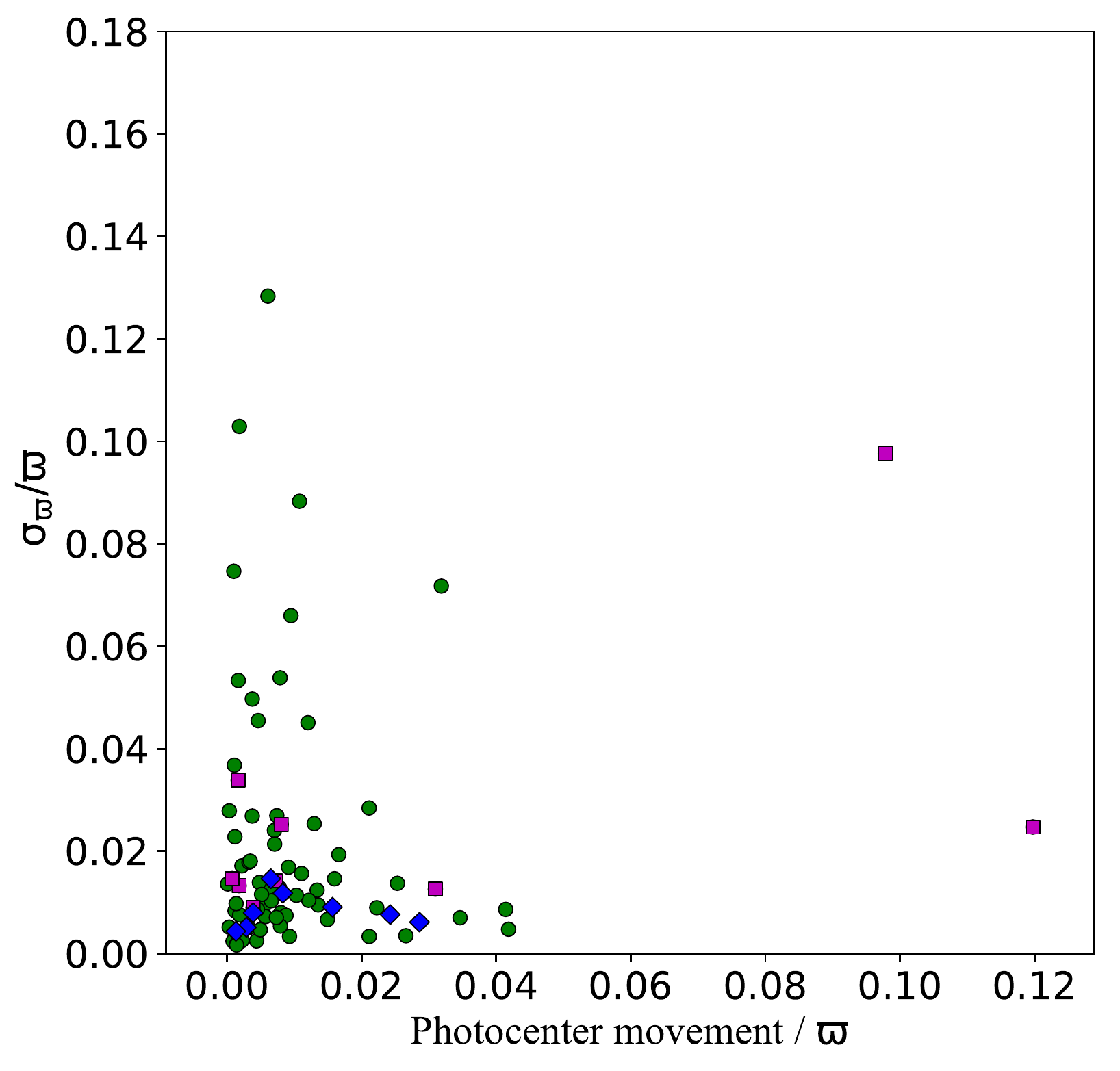}} 
\caption{Photocenter movement of eclipsing binaries in our sample expressed as a fraction of {\it Gaia} DR2 parallax {\it vs.} the fractional uncertainty of the {\it Gaia} parallax.
The meanings of the symbols are the same as in Fig.~\ref{fig:gaia_frac}.}
\label{fig:gaia_photcent}
\end{figure}

Of special interest is the knowledge of how binarity will impact {\it Gaia} parallaxes, and the determination of the global zero-point shift. In order to quantify this, we calculated the photocenter movement for our $\sim$80 eclipsing binaries and compared them with the uncertainties on {\it Gaia} DR2 parallaxes -- see Fig.~\ref{fig:gaia_photcent}. For the overwhelming majority of the systems the photocenter movement is only a very small fraction of the parallax itself. Only in two systems could {\it Gaia} DR2 probably recognize the photocenter movement (EPIC 211409263 and V380 Cyg). Thus we can consider the {\it Gaia} DR2 parallaxes of the eclipsing binaries in our sample as practically unaffected by binarity.

\acknowledgments
We dedicate this paper to our colleague Zbigniew ("Zibi") Ko{\l}aczkowski who passed away much too early.

%\textcolor{red}
{We thank the anonymous referee for valuable comments and remarks on the manuscript.}
The research leading to these results  has received
funding from the European Research Council (ERC) under the European
Union's Horizon 2020 research and innovation program (grant agreement
No. 695099).

We are grateful for financial support from Polish National Science
Center grant MAESTRO UMO-2017/26/A/ST9/00446. Support from the BASAL Centro
de Astrof{\'i}sica y Tecnolog{\'i}as Afines (CATA) grant AFB-170002, the
Millenium Institute of Astrophysics (MAS) of the Iniciativa Cientifica
Milenio del Ministerio de Economia, Fomento y Turismo de Chile, project
IC120009 and the IdP II 2015 0002 64 grant of the Polish Ministry of
Science and Higher Education is also acknowledged. We are also thankful
to the staff in La Silla Observatory (ESO) and Las Campanas Observatory
(Carnegie) for their excellent support.

This work has made use of data from the European Space Agency (ESA) mission
{\it Gaia} (\url{https://www.cosmos.esa.int/gaia}), processed by the {\it Gaia}
Data Processing and Analysis Consortium (DPAC,
\url{https://www.cosmos.esa.int/web/gaia/dpac/consortium}). Funding for the DPAC
has been provided by national institutions, in particular, the institutions
participating in the {\it Gaia} Multilateral Agreement.

This research has made extensive use of the excellent astronomical
SIMBAD database and of the VizieR catalog access tool, operated at CDS,
Strasbourg, France, and made also use of NASA's Astrophysics Data System
Bibliographic Services (ADS).

This publication makes use of data products from the Two Micron All
Sky Survey, which is a joint project of the University of Massachusetts
and the Infrared Processing and Analysis Center/California Institute of
Technology, funded by the National Aeronautics and Space Administration
and the National Science Foundation.

{}    

%\begin{appendix}
%\end{appendix}

\end{document}